\documentclass{aip-cp}

\usepackage[numbers]{natbib}
\usepackage{rotating}
\usepackage{graphicx}

\usepackage{bm}
\usepackage{multirow}
\usepackage{subfigure}
\begin{document}

% Title portion
%
\title{Constraining the Hadron-Quark Phase Transition Chemical Potential via Astronomical Observation}

\author[aff1,aff2]{Zhan Bai\noteref{note1}}
%\eaddress{baizhan@pku.edu.cn}

\author[aff1,aff2,aff3]{Yu-xin Liu\noteref{note2}}  %\corref{cor1}}
%\eaddress{yxliu@pku.edu.cn}

\affil[aff1]{Department of Physics and State Key Laboratory of Nuclear Physics and Technology, Peking University, Beijing 100871, China}
\affil[aff2]{Collaborative Innovation Center of Quantum Matter, Beijing 100871, China}
\affil[aff3]{Center for High Energy Physics, Peking University, Beijing 100871, China}

%\corresp[cor1]{Corresponding author: yxliu@pku.edu.cn}

\authornote[note1]{Speaker. Email: baizhan@pku.edu.cn}
\authornote[note2]{Correspondence could be addressed. Email: yxliu@pku.edu.cn}

\maketitle

\begin{abstract}
We investigate the chemical potential and baryon number density of the hadron-quark phase transition in neutron star matter.
The hadron matter is described with the relativistic mean field theory,
and the quark matter is described with the Dyson-Schwinger equation approach of QCD.
In order to study the first-order phase transition,
we develop the sound speed interpolation scheme to construct the equation of state in the middle density region
where the hadron phase and quark phase coexist.
The phase transition chemical potential is constrained with the maximum mass, the tidal deformability
and the radius of neutrons stars.
And the most probable value of the phase transition chemical potential is found.
\end{abstract}

% Head 1
\section{\label{sec:Intro}INTRODUCTION}

Researches on the phase transitions of strong interaction matter (QCD phase transitions. Or simply, hadron--quark phase transition) have attracted great attentions in recent years (for example, see Refs.~\cite{Roberts:200012,Gyulassy:2005NPA,Aoki:2006NAT,Blaizot:2007Review,Wambach:2009Review,Fukushima:201112Reviews,Qin:2011PRL,Fischer:2018,Philipsen:2013,Ding:Review}) since they may shed light on revealing the process and the nature of the early universe matter evolution.
The matter covers naturally that at high temperature and low (even vanishing) density (small baryon chemical potential), at intermediate temperature and density (warm dense matter), and that at low (even zero) temperature and high density (cold dense matter).
Since the phase transitions take place in the non-perturbative QCD energy scale,
one must take the non-perturbative QCD approaches to carry out the investigations on theoretical side.
On experimental side, one could generate the matter at high temperature and low density even the warm dense matter with the relativistic heavy ion collisions (RHICs) in laboratory and study the corresponding phase transitions.
However, one could not get the very cold dense matter in laboratory.
Therefore, one must take aid of astronomical observations, especially those for compact pulsars (see, e.g.,
Refs.~\cite{Glendenning:2000,Lattimer:2004,Weber:2005}).

Neutron star is a wonderful laboratory for the study of cold dense QCD matter,
since neutron star is one of the most compact objects in the universe,
and it is believed that there exists hadron-quark phase transition in its center region.
The neutron stars with a hadron mantle and a quark core is usually referred to as ``hybrid stars''.
One of the important properties of neutron stars that can be observed on earth is their mass.
Several years ago, two neutron stars with large mass (around $2M_{\odot}$) were
observed~\cite{Demorest:2010Nature,Antoniadis:2013Science},
which indicates that the equation of state (EoS) of the neutron star matter should be stiff.

Another important observation of neutron star is the gravitational wave.
The detection of gravitational wave GW170817 is the first observation of gravitational wave of binary neutron star merger~\cite{Abbott:2017APJL,Abbott:2017PRL},
and it marks a new era of multi-messenger astronomy.
This observation provides information about the tidal deformability of neutron stars.
According to Ref.~\cite{Abbott:2017PRL}, the tidal deformability of a $1.4M_{\odot}$ neutron star should be $\Lambda_{1.4}<800$ with a credence level of $90\%$,
which indicates that the EoS should be soft.
Therefore, the observation of the mass and tidal deformability together provide strong constraints on the EoS of neutron star matter.
Besides, it has been proposed that the gravitational oscillation frequency of the newly born neutron star
(soon after the supernova explosion) can also be a signal for the hadron--quark phase transition in the star matter for more than ten years~\cite{Liu:2008PRL}.

Although the astronomical observation has provided a lot of information for the EoS of cold dense matter,
the theoretical study is still limited.
In order to take into consideration of the hadron-quark phase transition,
the best way is making use of a unified lagrangian which can describe the hadron matter at low density
and so do the quark matter at high density.
However, for now we do not have such an approach.
The most common way is then to describe the hadron matter and the quark matter separately
via respective approach,
and combine them together to get the complete EoS of the matter in the whole density region
with constructions.

One of the most commonly used method for building the complete EoS of the matter involving a first-order phase transition is the Gibbs construction (e.g., Refs.~\cite{Glendenning:1992PRD,Glendenning:2000,Burgio:2002PRC,Carroll:2009PRC,Weissenborn:2011ApJL,Fischer:2011ApJS,Schulze:2011PRC}),
but it has its limitations.
For example, it assumes that the hadron model and the quark model are accurate at all densities.
In fact, as we know,
although the hadron model is accurate at saturation density because there are plenty of experimental data,
the hadrons cannot be regarded as point particles in the phase transition region,
and different models give results deviating from each other greatly in the phase transition region~\cite{Xiao:2009PRL,Oertel:2017RMP}.
This means that the hadron model becomes unreliable in the phase transition region.
Similarly, the quark model is usually only accurate in high density region where perturbative QCD can be applied,
in the phase transition density, the quark matter becomes non-perturbative and the quark model loses its accuracy.

To avoid these problems, the 3-window construction model has been
proposed~\cite{Masuda:2013PTEP,Masuda:2013ApJ,Kojo:2015PRD,Kojo:2016EPJA,Masuda:2016EPJA}.
In 3-window construction, the EoS in the phase transition region is derived by smooth interpolation.
However, the 3-window construction cannot calculate the detail of the phase transition consistently,
for example, the chemical potential or the baryon density at which the phase transition occurs,
i.e., the region for both the hadron phase and the quark phase to coexist, is assigned artificially,
since theoretical studies have not yet reached a common idea for the coexistence region of the phase transition.

It is well known that a typical first-order phase transition involves a phase coexistence region,
and the EoS is not smooth at the two boundaries where the phase coexistence region begins, disappears, respectively.
Consequently, the speed of sound, which is the derivative of the EoS, is discontinuous at the boundaries.
Therefore, if we make use of the speed of sound to do the interpolation,
we can study the phase transition with 3-window construction.

In this talk, we represent a scheme to determine the hadron--quark phase transition region
with astronomical observations. In our approach, we construct the speed of sound in the matter
(the EoS of the matter)
with a large set of values of the beginning and ending chemical potentials of the phase transition region.
Then various astronomical observables can be calculated with the obtained EoS of the star matter.
With the maximum mass, the radius and the tidal deformability of neutron stars being taken
as the calibration quantities to exclude the inappropriate constructions,
the baryon chemical potential range for the hadron-quark phase transition to happen can be fixed.

The remainder of this representation is organized as follows.
After this introduction, we review very briefly the studies on the QCD phase transitions
and show that the chemical potential region of the hadron-quark phase transition
has not yet been settled well on theoretical side at first.
Then we describe the hadron model and the quark model we take.
Thereafter we describe our scheme to construct the sound speed with interpolation,
and show how the range of the baryon chemical potential
in which the hadron-quark phase transition takes place is constrained by astronomical observations.
Next we illustrate our numerical results.
Finally we give a summary and some remarks.

\section{\label{sec:RevQCDPTs} QCD Phase transitions and the Phase Diagram}

Since the QCD phase transitions (or simply, hadron--quark phase transition) is one of the most significant topic in current physics, and the fundamental approach to carry out the research should be the non-perurbative QCD.
For instance, the lattice QCD simulations have given the pseudo-critical temperature for the chiral crossover (transition) at vanishing chemical potential, the phase diagram at low chemical potential and the feature in external magnetic filed (see, e.g., Refs.~\cite{Aoki:2006NAT,Philipsen:2013,Philipsen:201014PRL,Borsanyi:2010JHEP,DElia:20101PRD,Gupta:2011Sci,Liu:2011PRD,Gavai:2011PLB,Bazavov:2012PRD,Bali:2012JHEP,Ejiri:2013PRL,Bali:2014JHEP,Bonati:2014PRD,Levkova:2014PRL,Bornyakov:2014PRD,Ding:20167,Ding:2018}).  The Dyson-Schwinger (DS) equation approach~\cite{Roberts:DSE1} has provided the phase diagram, some of the variation behaviors of the thermodynamical quantities and the viscosities, even the effect of the interface between the hadron phase and the quark phase (see, e.g., Refs.~\cite{Roberts:200012,Qin:2011PRL,Fischer:2018,Chen:2008PRD,QCDPT-DSE21,Gao:2016PRDa,Zong:2015AP,Gao:20168PRD}).
The functional renormalization group (FRG) approach has also given the characteristics of the phase diagram, the temperature dependence of some transport properties and the baryon number fluctuations (e.g.,  Refs.~\cite{Pawlowski:2013,Pawlowski:2015PRL,Pawlowski:20156,Fu:201567PRD}).
Meanwhile general analysis and effective QCD models have also played the important role and made significant progress (see, e.g., Refs.~\cite{Blaizot:2007Review,Wambach:2009Review,Fukushima:201112Reviews,Rehberg:1996PRC,Ratti:2006PRD,Stephanov:2007fk,McLerran:2007NPA,Schaefer:2007PRD,Zhao:2008EPJC,Liu:Phenomodel,Endrodi:2013JHEP,Jiang:2013PRD,Ke:2014PRD}).

It has been known that the fundamental approach to investigate the QCD phase transitions should be the non-perturbative QCD.
To identify the phase boundary of a phase transition, one takes conventionally the order parameters and the effective thermodynamical potential as the criteria.
For the QCD phase transitions, especially the chiral phase transition which generates almost all the observable mass of the observable matter, one usually implement the (chiral) quark condensate as the order parameter.
However one could not get the effective thermodynamical potential in the fully non-perturbative QCD approach.
One develops then the chiral susceptibility as the criterion~\cite{Qin:2011PRL,Gao:2016PRDa,Zong:2015AP,Zhao:2008EPJC}.
(for the details about the equivalence between the susceptibility criterion and the thermodynamical potential criterion, please see Ref.~\cite{Gao:2016PRDa,Zhao:2008EPJC})

The (chiral) susceptibility is conventionally the responsibility (more explicitly, the derivative) of the order parameter  with respect to the control variables (for instance, the temperature, the chemical potential, the magnetic field strength, current mass, etc.). For multi-flavor system, it is the modula of the determinant of the concerned flavors' susceptibilities
(for details,  please see Ref.~\cite{Jiang:2013PRD}) .

\begin{figure*}[!htb]
\vspace*{-2mm}
\center
\includegraphics[width=0.40\textwidth]{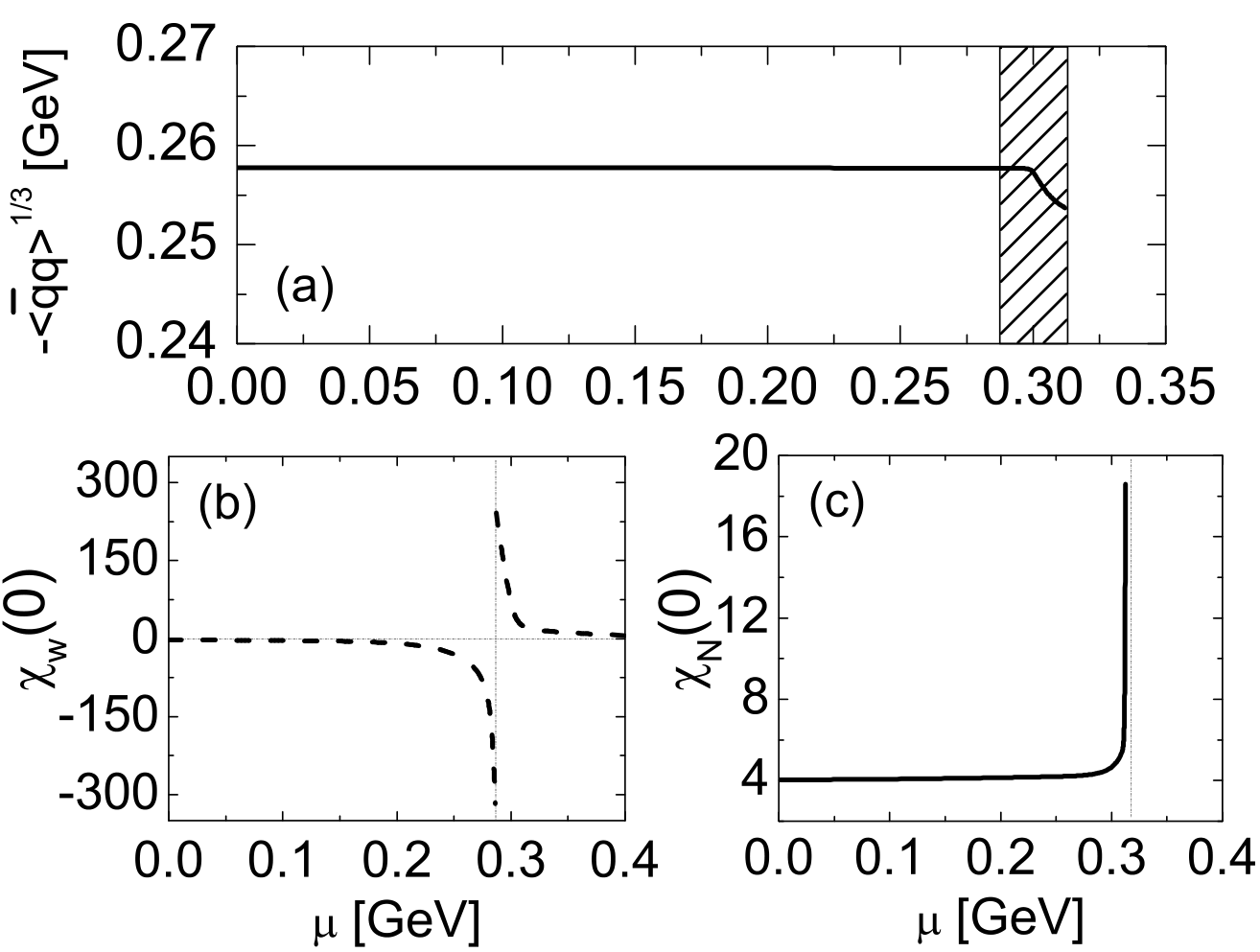} \hspace*{5mm}
\includegraphics[width=0.40\textwidth]{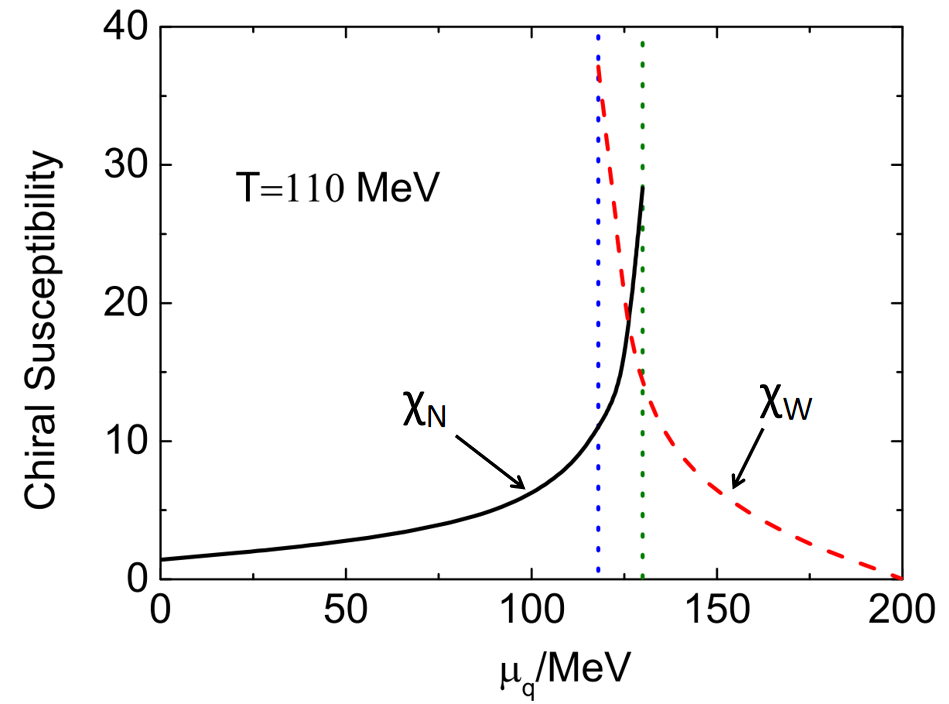}
\vspace*{-2mm}
\caption{
{\it Left panel:} Calculated quark condensate and chiral susceptibility as a function of quark chemical potential at $T=0$.
The result is obtained with the Ball-Chiu ansatz for the quark-gluon interaction vertex~\cite{Ball:1980PRD,Maris:2001PRC} and the simplified Maris--Tandy model~\cite{MTmodel-Gluon}
with parameters $D=1.0$GeV$^2$, $\omega=0.5$GeV for the gluon propagator (taken from the first of Ref.~\cite{Qin:2011PRL})
{\it Right panel:} Similar to the left panel but for the case $T=110\,$MeV and obtained from the fully dressed quark-gluon vertex~\cite{CLR-vertex}
and the Qin-Chang model~\cite{QCmodel-Gluon} with parameters $D=0.28$GeV$^2$ and $\omega=0.5$GeV
for the gluon propagator (re-plotted from Ref.~\cite{Gao:2016PRDa}).
}
\vspace*{-3mm}
\label{fig:ChiralSus}
\end{figure*}

With the DS equation approach (a brief description is given in next Section. For details, please see Refs.~\cite{Roberts:DSE1,Roberts:200012,Fischer:2018}),
one can calculate the quark condensate and chiral susceptibility with various algorithms and different models for the propagator and the quark-gluon interaction vertex.
Two examples of the obtained results are shown in Fig.~\ref{fig:ChiralSus}.
As can be seen from the figure,
in lower temperature region, the susceptibility of the Nambu (the dynamical chiral symmetry breaking, DCSB) phase (the one with observable dynamical mass)
diverges at the states different from those for that of the Wigner (dynamical chiral symmetric, DCS) phase (the one with zero mass or the one corresponding to the current (or simply the bare) mass) to diverge.
This means that the phase transition is in first-order.

The region between the states for the susceptibilities of the two phases to diverge individually is the phase coexistence region.
In the crossover region, the susceptibility is a smooth function involving a peak,
{\it i.e.}, the susceptibility of the DCSB phase links with (in fact, changes to)  that of the DCS phase not only continuously but also smoothly.
While, in the second-order transition region, the susceptibility of the DCSB phase diverges
at the same location as that for the susceptibility of the DCS phase to diverge.
The characteristic for the susceptibilities of the two phases to diverge at the same location defines thus the state which separates the crossover region from the first-order transition region, namely, the critical endpoint (CEP). Therefore the chiral susceptibility criterion can not only give the chiral phase transition boundaries
but also localize the position of the CEP.

\begin{figure*}[!htb]
\vspace*{-1mm}
\center
\includegraphics[width=0.45\textwidth]{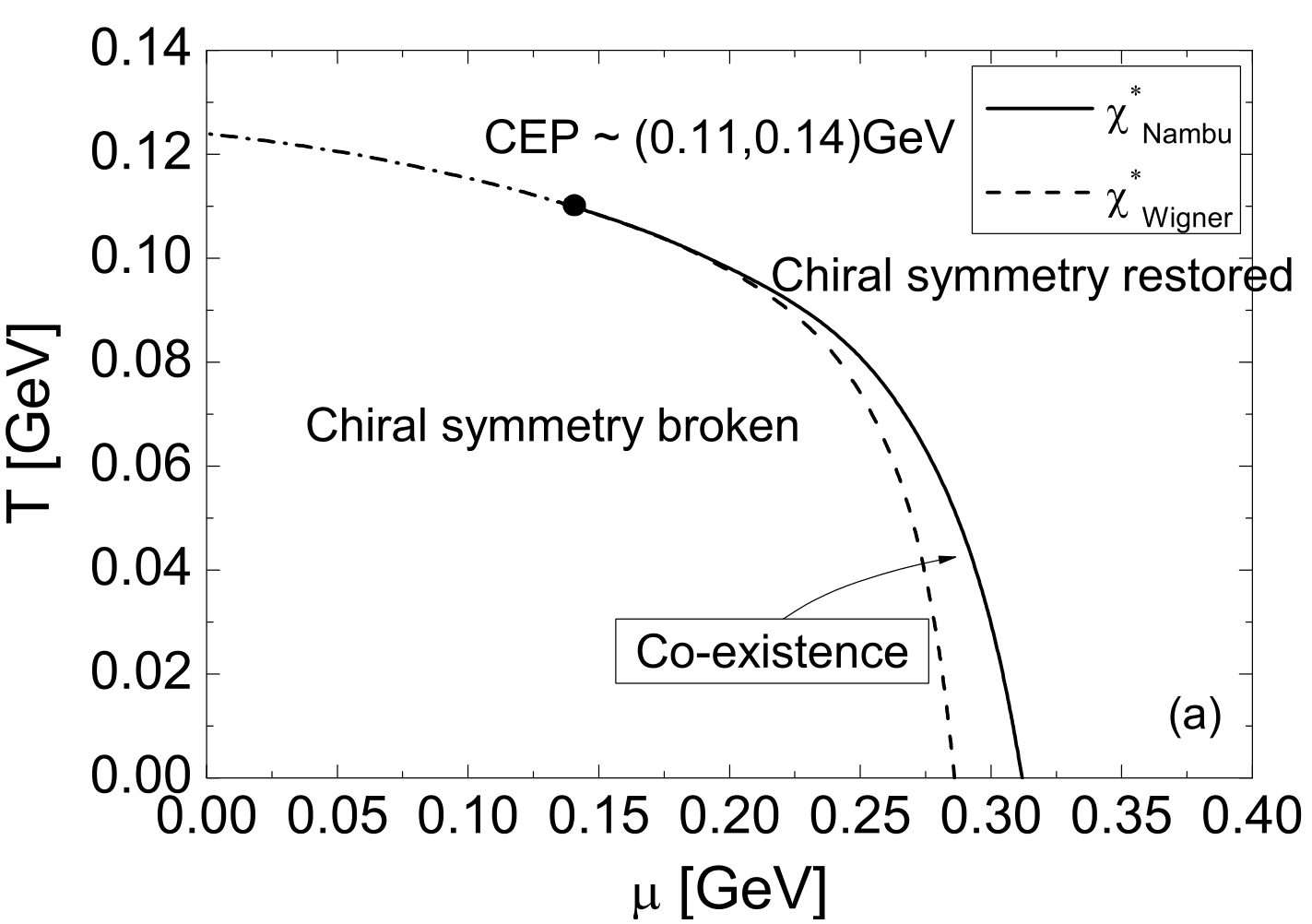} \hspace*{4mm}
\includegraphics[width=0.40\textwidth]{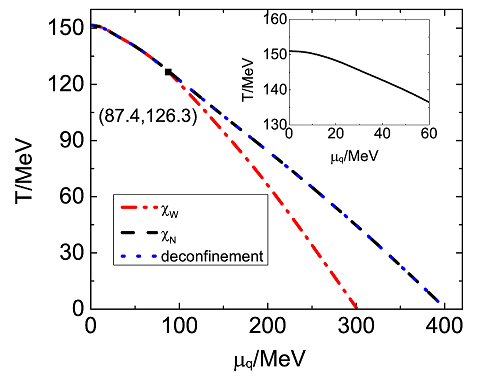}
\vspace*{-2mm}
\caption{Calculated QCD phase diagrams in terms of temperature and quark chemical potential.
The parameters used in the left and the right panel is the same as those for the left, right panel of Fig.~\ref{fig:ChiralSus}, and also taken from Ref.~\cite{Qin:2011PRL}, \cite{Gao:2016PRDa}, respectively.}
\vspace*{-2mm}
\label{fig:QCDPD-DSE}
\end{figure*}

Using the DS equation approach, one can calculate the QCD phase digram.
Two examples of the obtained results are displayed in Fig.~\ref{fig:QCDPD-DSE}.

The results shown in Fig.~\ref{fig:QCDPD-DSE} and other DS equation calculations indicate that the pseudo-critical temperature of the chiral crossover (transition) and the temperature of the CEP have been settled down well.
However different calculations give quite distinct results for the baryon chemical potential at the CEP and for the chemical potentials identifying the phase coexistence region at vanishing temperature (see, e.g., Refs.~\cite{Qin:2011PRL,Fischer:2018,QCDPT-DSE21,Gao:2016PRDa}).
Inferring the recent lattice QCD simulation result given in Ref.~\cite{Ding:2018} one can get the similar conclusion.
Concerning the results given in effective QCD models one can know that the CEP may locates at much higher baryon chemical potential and the phase coexistence region at zero temperature takes much higher baryon chemical potentials too.
Since theoretical investigations can not give solid results (consistent with each other) of the baryon chemical potentials for the hadron phase and the quark phase to coexist and the RHIC can not (at least at present stage) provide such a knowledge either, one should take advantage of the astronomical observations.
In this talk we would like very much to represent a scheme to constrain the baryon chemical potentials of the hadron-quark phase coexistence region with astronomical observations.

% Head 2
\section{\label{sec:model} Hadron and Quark Models}
\subsection{\label{sec:RMF}Hadron Sector}

As mentioned above we intend to establish a scheme to constrain the boundaries of the hadron-quark phase transition region at vanishing temperature with astronomical observations, in which the speed of sound in the matter is taken as the signature quantity to label the phase transition.
It is well known that in order to calculate the speed of sound for compact star matter,
we need to calculate the EoS of the matter in different phases (hadron phase, or quark phase,
or the coexistence phase) at first.

For hadron matter, we adopt the TW-99 model of the relativistic mean field (RMF) theory~\cite{Typel:1999NPA}.
The Lagrangian of the TW-99 model for the hadron matter is
\begin{equation}\label{eqn:Lagrangian}
\mathcal{L}=\mathcal{L}_{B}^{} + \mathcal{L}_{\small \textrm{lep}}^{} + \mathcal{L}_{M}^{}
+ \mathcal{L}_{\small \textrm{int}}^{} \, ,
\end{equation}
where $\mathcal{L}_{B}^{}$ is the Lagrangian of free baryons.

In this work, we consider the baryon octet $p$,$n$,$\Lambda$,$\Sigma^{\pm,0}$ and $\Xi^{-,0}$ for baryons.
The corresponding Lagrangian reads
\begin{equation}
\mathcal{L}_{\scriptsize\textrm{B}}^{} = \sum_{\small\textrm{i}}{\bar{\Psi}_{i}^{}} (i\gamma_{\mu}^{} \partial^{\mu} - m_{i}^{}){\Psi_{i}^{}}\, .
\end{equation}

$\mathcal{L}_M$ is the Lagrangian of mesons,
\begin{eqnarray}
\mathcal{L}_{M}^{} =  \frac{1}{2}\left(\partial_{\mu}^{} \sigma \partial^{\mu} \sigma - m_{\sigma}^{2} \sigma^{2} \right) -\frac{1}{4} \omega_{\mu\nu}^{} \omega^{\mu\nu}
- \frac{1}{2}m_{\omega}^{2} \omega_{\mu}\omega^{\mu}
 -\frac{1}{4}\boldsymbol{\rho}_{\mu\nu}^{} \boldsymbol{\rho}^{\mu\nu} -\frac{1}{2}m_{\rho}^{2} \boldsymbol{\rho}_{\mu} \boldsymbol{\rho}^{\mu} \, ,
\end{eqnarray}
where $\sigma$, $\omega_{\mu}^{}$, and $\boldsymbol{\rho}_{\mu}^{}$ are the
isoscalar-scalar, isoscalar-vector and isovector-vector meson field, respectively.
$\omega_{\mu\nu}^{}=\partial_{\mu}\omega_{\nu} - \partial_{\nu}\omega_{\mu}$,
$\boldsymbol{\rho}_{\mu\nu}^{} =\partial_{\mu}\boldsymbol{\rho}_{\nu} - \partial_{\nu}\boldsymbol{\rho}_{\mu}\,$.

The $\mathcal{L}_{\small\textrm{int}}$ in Eq.~(\ref{eqn:Lagrangian}) is the Lagrangian
 describing the interactions between baryons which are realized by exchanging the mesons:
\begin{eqnarray}
\mathcal{L}_{\small \textrm{int}} =  \sum_{B} g_{\sigma B}^{} {\bar{\Psi}_{B}^{}} \sigma {\Psi_{B}^{}}
-g_{\omega B}^{} {\bar{\Psi}_{B}^{}} \gamma_{\mu}^{} \omega^{\mu} {\Psi_{B}^{}} - g_{\rho B}^{} {\bar{\Psi}_{B}^{}} \gamma_{\mu}^{} \boldsymbol{\tau}_{B}^{} \cdot \boldsymbol{\rho}^{\mu} {\Psi_{B}^{}} \, ,
\end{eqnarray}
where $g_{iB}^{}$ for $i=\sigma$, $\omega$, $\rho$ are the coupling strength parameters between baryons and mesons, which depend on the baryon density.

For nucleons, the coupling constants are
\begin{equation}
{g_{iN}^{}}(\rho_{B}^{}) = {g_{iN}^{}} (\rho_{\small\textrm{sat}}^{}) f_{i}^{} (x),  \qquad \quad \textrm{for} \quad i=\sigma,\omega, \rho,
\end{equation}
where $\rho_{B}^{}$ is the baryon density, $\rho_{\small\textrm{sat}}^{}$ is the saturation density of nuclear matter  and
$x= {\rho_{B}^{}}/{\rho_{\small\textrm{sat}}^{}}$.
The density function can be written as~\cite{Typel:1999NPA}:
\begin{eqnarray}
f_{i}^{}(x) & = & a_{i}^{} \frac{1+b_{i}^{} (x + d_{i}^{})^{2}}{1 + c_{i}^{}(x+d_{i}^{})^{2}}, \qquad \textrm{for} \qquad i=\sigma,\omega \, , \\
f_{\rho}^{}(x)&  = &\textrm{exp}\left[-a_{\rho}^{} (x-1) \right] \, ,
\end{eqnarray}
where the value of parameters $a_{i}^{}$, $b_{i}^{}$, $c_{i}^{}$, $d_{i}^{}$ and $g_{iN}^{}(\rho_{\small\textrm{sat}}^{})$
and $m_{i}^{}$ are listed in Table~\ref{tab:gi}.

\begin{table}[htb]
\caption{Parameters of the mesons and their couplings (taken from Ref.~\cite{Typel:1999NPA}).}
\label{tab:gi}
\begin{tabular}{cccc}
\hline
Meson i & $\sigma$ & $\omega$ & $\rho$ \\
\hline
~$m_i$(MeV)~ & 550 & 783 & 763\\
$g_{iN}^{}(\rho_{\textrm{sat}}^{})$ & ~$10.72854$~ & ~$13.29015$~ & ~$7.32196$~ \\
$a_{i}^{}$ & ~$1.365469$~ & ~$1.402488$~ & ~$0.515$~ \\
$b_{i}^{}$ & ~$0.226061$~ & ~$0.172577$~ & {} \\
$c_{i}^{}$ & ~$0.409704$~ & ~$0.344293$~ & {} \\
$d_{i}^{}$ & ~$0.901995$~ & ~$0.983955$~ & {} \\
\hline
\end{tabular}
\end{table}

For hyperons, we represent them with the relation between the hyperon coupling and the nucleon coupling as:
$\chi_{\sigma}^{}=\frac{g_{\sigma Y}^{}}{g_{\sigma N}^{}}$,
$\chi_{\omega}^{}=\frac{g_{\omega Y}^{}}{g_{\omega N}^{}}$,
$\chi_{\rho}^{}=\frac{g_{\rho Y}^{}}{g_{\rho N}^{}}$.
On the basis of hypernuclei experimental data, we choose them as those in Refs.~\cite{Glendenning:2000,Dutra:2016PRC}:
$\chi_{\sigma}^{}=0.7$, $\chi_{\omega}^{}=\chi_{\rho}^{}=0.783$.

The $\mathcal{L}_{\small\textrm{lep}}^{}$ is the Lagrangian for leptons, which are treated as free fermions:
\begin{equation}
\mathcal{L}_{\small\textrm{lep}}^{}=\sum_{l}^{} {\bar{\Psi}_{l}^{}}(i\gamma_{\mu}^{} \partial^{\mu} - m_{l}^{}) {\Psi_{l}^{}} \, ,
\end{equation}
and we include only the electron and muon in this work.

The field equations can be derived by differentiating the Lagrangian.
Under RMF approximation, the system is assumed to be in the static, uniform ground state.
The partial derivatives of the mesons all vanish,
except that only the 0-component of the vector meson and the 3rd-component of the isovector meson survive and can be replaced with the corresponding expectation values.
The field equations of the mesons are then:
\begin{equation}\label{eqn:sigma}
m_{\sigma}^{2} \sigma =\sum_{B} g_{\sigma B}^{} \langle {\bar{\Psi}_{B}^{}} {\Psi_{B}^{}} \rangle \, ,
\end{equation}
\begin{equation}\label{eqn:omega}
m_{\omega}^{2} \omega_{0}^{} = \sum_{B} g_{\omega B}^{} \langle {\bar{\Psi}_{B}^{}} \gamma_{0}^{} {\Psi_{B}^{}} \rangle \, ,
\end{equation}
\begin{equation}\label{eqn:rho}
m_{\rho}^{2} \rho_{03}^{} = \sum_{B} g_{\rho B}^{} \langle {\bar{\Psi}_{B}^{}} \gamma_{0}^{} \tau_{3B}^{} {\Psi_{B}^{}} \rangle \, ,
\end{equation}
where $\tau_{3B}^{}$ is the 3rd-component of the isospin of baryon $B$.

The equation of motion (EoM) of the baryon is:
\begin{equation}\label{EOM}
\left[\gamma^{\mu} (i\partial_{\mu}^{} - \Sigma_{\mu}^{}) - (m_{B}^{} - g_{\sigma B}^{} \sigma) \right] {\Psi_{B}^{}} = 0 \, ,
\end{equation}
where
\begin{equation}
\Sigma_{\mu}^{} = g_{\omega B}^{} \omega_{\mu} + g_{\rho B}^{} \boldsymbol{\tau}_{B}^{} \cdot \boldsymbol{\rho}_{\mu}^{} + \Sigma_{\mu}^{\scriptsize\textrm{R}}.
\end{equation}
The $\Sigma_{\mu}^{\scriptsize \textrm{R}}$ is called the ``rearrange'' term,
which appears because of the density-dependence of the coupling constant, and reads
\begin{eqnarray}\label{eqn:rearrange}
\Sigma_{\mu}^{\scriptsize\textrm{R}} =  \frac{j_{\mu}^{}}{\rho} \bigg(\frac{\partial g_{\omega B}^{}} {\partial\rho}\bar{\Psi}_{B}^{} \gamma^{\nu} \Psi_{B}^{} \omega_{\nu}^{}
	 + \frac{\partial g_{\rho B}^{}}{\partial\rho}\bar{\Psi}_{B}^{} \gamma^{\nu} \boldsymbol{\tau}_{B}^{} \cdot \boldsymbol{\rho}_{\nu}^{} \Psi_{B}^{}
		-\frac{\partial g_{\sigma B}^{}}{\partial\rho}\bar{\Psi}_{B}^{} \Psi_{B}^{} \sigma \bigg) \, ,
\end{eqnarray}
where $j_{\mu}^{} = \bar{\Psi}_{B}^{} \gamma_{\mu} \Psi_{B}^{}$ is the baryon current.

Under the EoM of Eq.~(\ref{EOM}),
the baryons behave as quasi-particles with effective mass
\begin{equation}\label{eqn:mstar}
m_{B}^{\ast} = m_{B}^{} - g_{\sigma B}^{} \sigma \, ,
\end{equation}
and effective chemical potential:
\begin{equation}\label{eqn:mustar}
\mu_{B}^{\ast} = \mu_{B}^{} - g_{\omega B}^{} \omega_{0}^{} - g_{\rho B}^{} \tau_{3B}^{} \rho_{03}^{} -\Sigma_{\mu}^{\scriptsize \textrm{R}} \, .
\end{equation}

One can then get the baryon (number) density:
\begin{equation}\label{eqn:ni}
\rho_{B}^{} \equiv \langle {\bar{\Psi}_{B}^{}} \gamma^{0} {\Psi_{B}^{}} \rangle =\gamma_{B}^{}\int\frac{\textrm{d}^3k}{(2\pi)^3} = \gamma_{B}^{}\frac{k_{FB}^{3}}{6\pi^{2}} \, ,
\end{equation}
where $k_{FB}^{} =\sqrt{\mu_{B}^{\ast 2} - m_{B}^{\ast 2}}$ is the Fermi momentum of the particle,
$\gamma_{B}^{} =2$ is the spin degeneracy.
And the scalar density is:
\begin{equation}\label{eqn:nis}
	\rho_{B}^{s} \equiv  \langle {\bar{\Psi}_{B}^{}} {\Psi_{B}^{}} \rangle =\gamma_{B}^{}\int\frac{\textrm{d}^3k}{(2\pi)^{3}}\frac{m_{B}^{\ast}}{\sqrt{k^{2} + m_{B}^{\ast 2}}}
	=\gamma_{B}^{}\frac{m_{B}^{\ast}}{4\pi^{2}}\bigg[k_{FB}^{} \mu_{i}^{\ast} - m_{B}^{\ast 2} \textrm{ln}\bigg(\frac{k_{FB}^{} + \mu_{B}^{\ast}}{m_{B}^{\ast}}\bigg)\bigg] \, .
\end{equation}

The density of a kind of leptons is the same as that for baryons,
except that the effective mass and the effective chemical potential should be replaced with the corresponding mass and chemical potential of the lepton:
\begin{equation}\label{eqn:nl}
\rho_{l}^{} = \frac{k_{Fl}^{3}}{3\pi^{2}} \, ,
\end{equation}
where $k_{Fl}^{2} = \mu_{l}^{2} - m_{l}^{2}$ for $l=e^{-},\, \mu^{-}$.

The matter in the star composed of hadrons should be in $\beta$-equilibrium.
Since there are two conservative charge numbers: the baryon number and the electric charge number,
all the chemical potential can be expressed with the neutron chemical potential and the electron chemical potential:
\begin{equation}\label{eqn:beta}
\mu_{i}^{} = B \mu_{n}^{} - Q \mu_{e}^{} \, ,
\end{equation}
where $B$ and $Q$ is the baryon number, electric charge number for the particle $i$, respectively.

Then, combining Eqs.~(\ref{eqn:sigma}), (\ref{eqn:omega}), (\ref{eqn:rho}), (\ref{eqn:rearrange}), (\ref{eqn:mstar}),
(\ref{eqn:mustar}), (\ref{eqn:ni}), (\ref{eqn:nis}), (\ref{eqn:nl}) and (\ref{eqn:beta}),
together with the charge neutral condition:
\begin{equation}
\rho_{p}^{} + \rho_{\Sigma^{+}}^{}  = \rho_{e}^{} + \rho_{\mu^{-}}^{} + \rho_{\Sigma^{-}}^{}
+\rho_{\Xi^{-}}^{} \, ,
\end{equation}
one can determine the ingredients and the properties of the hadron matter with any given baryon density $\rho_{B}^{}\,$.

The EoS of the hadron matter can be calculated from the energy-momentum tensor:
\begin{equation}
T^{\mu\nu} = \sum_{\phi_i} \frac{\partial\mathcal{L}}{\partial(\partial_{\mu}\phi_{i}^{})}\partial^{\nu}\phi_{i}^{} - g^{\mu\nu}\mathcal{L}.
\end{equation}
The energy density $\varepsilon$ is:
\begin{equation}
\varepsilon=\langle T^{00}\rangle =\sum_{i=B,l}\varepsilon_{i}^{} +\frac{1}{2}m_{\sigma}^{2} \sigma^{2} +\frac{1}{2}m_{\omega}^{2} \omega_{0}^{2} + \frac{1}{2}m_{\rho}^{2} \rho_{03}^{2} \, ,
\end{equation}
where the contribution of the baryon $B$ to the energy density is:
\begin{equation}
\varepsilon_{B}^{}  = \gamma_{B}^{}\int\frac{\textrm{d}^3k}{(2\pi)^3}\sqrt{k^{2} +m_{B}^{\ast 2}}
	= \! \gamma_{B}^{}\frac{1}{4\pi^{2}} \Big[ 2\mu_{B}^{\ast 3} k_{FB}^{} \! - \! m_{B}^{\ast 2} \mu_{B}^{\ast} k_{FB}^{}
\! - \! m_{B}^{\ast 4}\textrm{ln}\Big(\! \frac{\mu_{B}^{\ast} \! + \! k_{FB}^{}}{m_{B}^{\ast}} \! \Big) \Big].
\end{equation}

The contribution of the leptons to the energy density can be written in the similar form as baryons with a spin degeneracy parameter $\gamma_{l}^{} = 2 $,
except that the effective mass and effective chemical potential should be replaced with those of the leptons, respectively.

As for the pressure of the system, we can determine that with the general formula:
\begin{equation}
P =\sum_{i} \mu_{i}^{} \rho_{i}^{} - \varepsilon \, .
\end{equation}

\begin{figure*}[!htb]
\vspace*{-2mm}
\center
\includegraphics[width=0.50\textwidth]{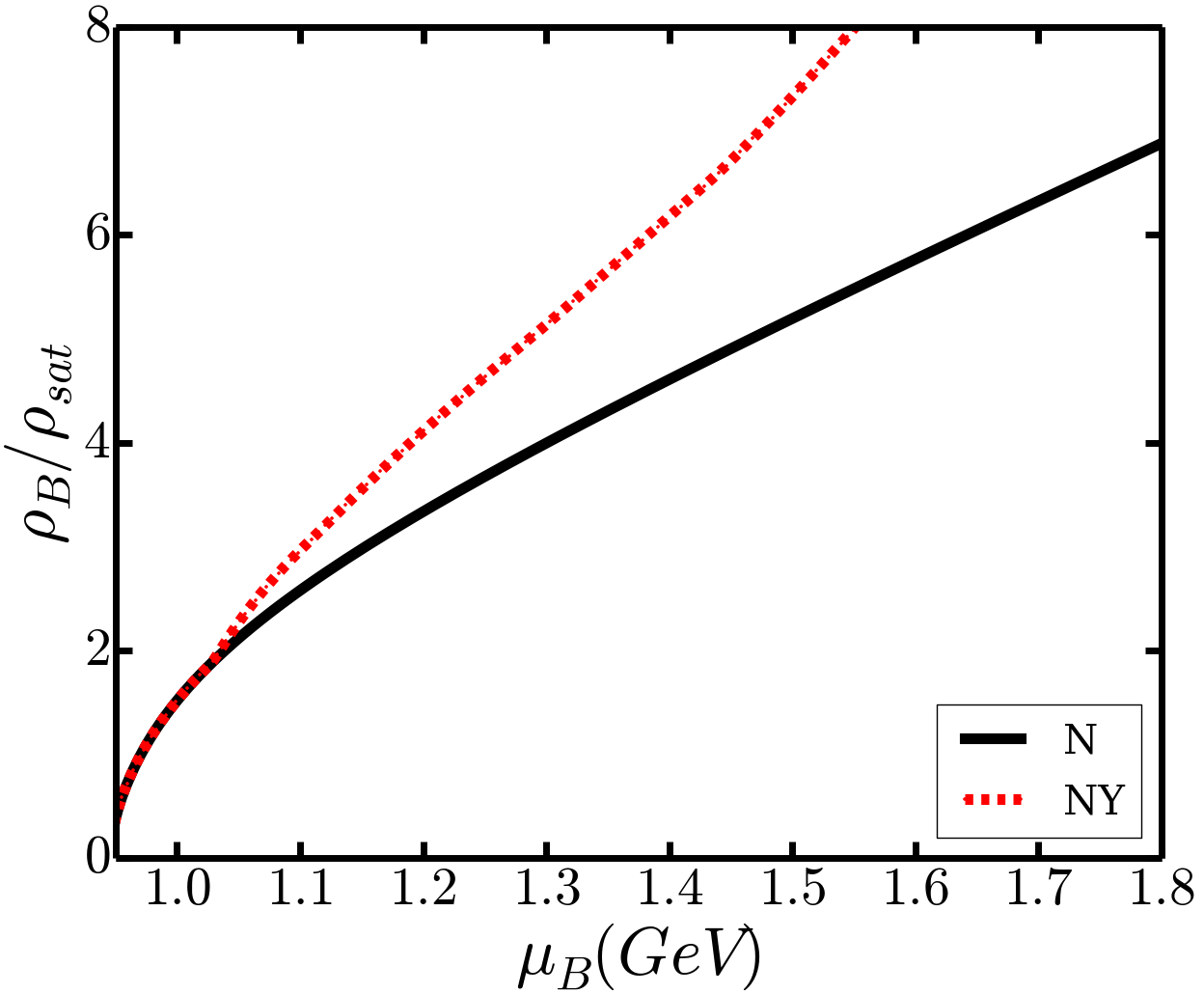}
\vspace*{-3mm}
\caption{Calculated relation between the baryon number density and the baryon chemical potential
via the TW-99 model in the RMF theory.
The (black) solid line and the (red) dashed line correspond to the result without and with hyperons, respectively.}
\label{fig:mu_n_H}
\end{figure*}

The calculated baryon number density as a function of baryon chemical potential is shown in Fig.~\ref{fig:mu_n_H}.
The black solid line and the red dashed line correspond to the result without and with hyperons, respectively.
The two lines are the same at low chemical potential, and they begin to split at $\mu=1.03\,\textrm{GeV}$.
This means that the hyperons begin to appear at this chemical potential,
which corresponds to $\rho_{B}^{}=1.90\,\rho_{\small\textrm{sat}}^{}$.

\subsection{\label{sec:quark} Quark Matter Sector}

For quark matter, we adopt the Dyson-Schwinger equation (DSE) approach (see, e.g., Refs.~\cite{Roberts:200012,Roberts:DSE1,CLR-vertex,QCmodel-Gluon,Roberts:DSE-BSE}).
The DSE approach is a continuum QCD approach
which includes both the confinement and the dynamical chiral symmetry breaking features simultaneously~\cite{McLerran:2007NPA},
and is successful in describing QCD phase transitions and hadron properties (see, e.g., Refs.~\cite{Roberts:DSE1,Roberts:200012,Qin:2011PRL,Fischer:2018,Chen:2008PRD,QCDPT-DSE21,Gao:2016PRDa,Gao:20168PRD,Maris:2001PRC,MTmodel-Gluon,CLR-vertex,QCmodel-Gluon,Roberts:DSE-BSE}).

The starting point of the DSE approach is the stability of the quark, the gluon and the ghost fields.
The truncated one can be written as the quark gap equation:
\begin{equation}
S(p;\mu)^{-1}=Z_{2}^{} [i\boldsymbol{\gamma} \cdot  \boldsymbol{p} + i\gamma_{4}^{} (p_{4}^{} + i\mu)+m_{q}^{}]+\Sigma(p;\mu),
\end{equation}
where $S(p;\mu)$ is the quark propagator, $\Sigma(p;\mu)$ is the renormalized self-energy of the quark:
\begin{eqnarray}
\Sigma(p;\mu)=Z_{1}^{} \int^{\Lambda}\frac{\textrm{d}^{4} q}{(2\pi)^{4}}
          g^{2}(\mu)D_{\rho\sigma}^{} (p-q;\mu)
		\times\frac{\lambda^{a}}{2} \gamma_{\rho} S(q;\mu) \Gamma_{\sigma}^{a}(q,p;\mu),
\end{eqnarray}
where $\int^{\Lambda}$ is the translationally regularized integral,
$\Lambda$ is the regularization mass-scale.
$g(\mu)$ is the strength of the coupling, $D_{\rho\sigma}^{}$ is the dressed gluon propagator,
$\Gamma_{\sigma}^{a}$ is the dressed quark-gluon interaction vertex,
$\lambda^{a}$ is the Gell-Mann matrix, and $m_{q}^{}$ is the current mass of the quark.
For simplicity, the current mass of $u$ and $d$ quark is now taken to be zero,
and the current mass of $s$ quark to be $115\;$MeV, by fitting the kaon mass in vacuum~\cite{Chen:2011PRD}.
$Z_{1,2}^{}$ are the renormalization constants.

At finite chemical potential,
the quark propagator can be decomposed according to the Lorentz structure as:
\begin{eqnarray}
S(p;\mu)^{-1}=i\boldsymbol{\gamma} \cdot  \boldsymbol{p} A(p^{2}, p \, u, \mu^{2})
+ B(p^{2},p\, u, \mu^{2})
		+i\gamma_{4}^{} (p_{4} + i\mu) C(p^{2},p\, u, \mu^{2}) \, ,
\end{eqnarray}
with $u=(\boldsymbol{0},i\mu)$.

At zero chemical potential, a commonly used ansatz for the dressed gluon propagator and the dressed quark-gluon interaction vertex is:
\begin{equation}
Z_{1}^{} g^{2}  D_{\rho\sigma}^{}(p-q)\Gamma_{\sigma}^{a}(q,p)
		=\mathcal{G}((p-q)^{2} )D_{\rho\sigma}^{\scriptsize \textrm{free}}(p-q) \frac{\lambda^a}{2}
             \Gamma_{\sigma}^{}(q,p) \, ,
\end{equation}
where
\begin{equation}
D_{\rho\sigma}^{\scriptsize \textrm{free}}(k\equiv p-q)=\frac{1}{k^{2}} \Big( \delta_{\rho\sigma}^{} - \frac{k_{\rho}^{} k_{\sigma}^{}}{k^{2}} \Big) \, ,
\end{equation}
$\mathcal{G}(k^{2})$ is the effective interaction introduced in the model,
and $\Gamma_{\sigma}^{}$ is the quark-gluon interaction vertex.

For the interaction part, we adopt the Gaussian type effective interaction (see, e.g., Refs.~\cite{Qin:2011PRL,Chen:2008PRD,MTmodel-Gluon,Chen:2011PRD,Alkofer:2002PRD,Chang:2005NPA}):
\begin{equation}
\frac{\mathcal{G}(k^{2})}{k^{2}}=\frac{4\pi^{2} D}{\omega^{6}}k^{2} \textrm{e}^{-k^{2}/\omega^{2}} \, ,
\end{equation}
where $D$ and $\omega$ are the parameters of the model.
In our present work we take $\omega=0.5\,$GeV and $D=1.0\,\textrm{ GeV}^2$ as the same as in many literatures.
In case of finite chemical potential, an exponential dependence of the ${\mathcal{G}}$ on the chemical potential was introduced in Ref.~\cite{Chen:2011PRD} as:
\begin{equation}\label{eqn:alpha}
\frac{\mathcal{G}(k^{2};\mu)}{k^{2}} = \frac{4\pi^{2} D}{\omega^{6}} \textrm{e}^{-\alpha\mu^{2}/\omega^{2}}k^{2} \textrm{e}^{-k^{2}/\omega^{2}} \, ,
\end{equation}
where $\alpha$ is the parameter controlling the rate for the quark matter to approach the asymptotic freedom.
It is evident that, when $\alpha=0$, it is the same as that at zero chemical potential;
when $\alpha=\infty$, the effective interaction is zero and corresponds to the case of MIT bag model.
We adopt such a model in our present calculations, and for simplicity, we take the same interaction for each flavor of the quarks.

For the quark--gluon interaction vertex, Ref.~\cite{Chen:2015PRD} has calculated the properties of quark matter with several different vertex models and shown that the vertex effect can be absorbed into the variation of the parameter $\alpha$.
We then in this work adopt only the rainbow approximation
\begin{equation}
\Gamma_{\sigma}^{}(q,p) = \gamma_{\sigma}^{} \, .
\end{equation}
In our previous work~\cite{Bai:2018PRD},
we have shown that when $\alpha=2$, the maximum mass of the hybrid star can be larger than $2M_{\odot}$.
We take then now $\alpha=2$.

With the above equations, we can get the quark propagator,
and derive the EoS of the quark matter in the same way as taken in Refs.~\cite{Qin:2011PRL,Chen:2008PRD,Gao:2016PRDa,Gao:20168PRD,Klahn:2010PRC}.

The number density of quarks as a function of its chemical potential is:
\begin{equation}
n_{q}^{}(\mu) = 6\int\frac{\textrm{d}^{3}p}{(2\pi)^{3}} f_{q}^{}(|\boldsymbol{p}|;\mu),
\end{equation}
where $f_{q}^{}$ is the distribution function and reads
\begin{equation}
f_{q}^{}(|\boldsymbol{p}|;\mu) = \frac{1}{4\pi}\int_{-\infty}^{\infty}\textrm{d}p_{4}^{} \textrm{tr}_{\textrm{D}}^{}[-\gamma_{4}^{} S_{q}^{}(p;\mu)] \, ,
\end{equation}
where the trace is for the spinor indices.

The pressure of each flavor of quark at zero temperature can be obtained by integrating the number density:
\begin{equation}
P_{q}^{}(\mu_{q}^{}) = P_{q}^{}(\mu_{q,0}^{}) + \int_{\mu_{q,0}^{}}^{\mu_{q}^{}} \textrm{d}\mu n_{q}^{}(\mu) \, .
\end{equation}

The total pressure of the quark matter is the sum of the pressure of each flavor of quark:
\begin{equation}
P_{Q}^{}(\mu_{u}^{},\mu_{d}^{},\mu_{s}^{}) = \sum_{q=u,d,s}\tilde{P}_{q}^{}(\mu_{q}^{}) - B_{\scriptsize \textrm{DS}}^{} \, ,
\end{equation}
\begin{equation}
\tilde{P}_{q}^{}(\mu_{q}^{})\equiv \int_{\mu_{q,0}^{}}^{\mu_{q}^{}}\textrm{d}\mu n_{q}^{}(\mu) \, ,
\end{equation}
\begin{equation}\label{eqn:B_DS}
B_{\scriptsize \textrm{DS}}^{} \equiv -\sum_{q=u,d,s} P_{q}^{}(\mu_{q,0}^{}) \, .
\end{equation}

Mathematically, the starting point of the integral $\mu_{q,0}^{}$ can be any value as long as we know the quark pressure $P_{q}(\mu_{q,0})$ at that chemical potential
(notice that $\mu_{q,0}$ is NOT the quark chemical potential corresponding to the onset of quark number density).
However, in practice, the quark pressure at arbitrary chemical potential cannot be fixed in prior.
%For the value of $B_{\scriptsize \textrm{DS}}^{}$,
%
%along the line of the discussion in Ref.~\cite{Chen:2016EPJA},
%
Therefore, in this work, we adopt the ``steepest-descent" approximation at zero chemical potential ($\mu_{B}=\mu_{u,d}=\mu_{s}=0$),
and implement $B_{\scriptsize \textrm{DS}}^{}
=90\textrm{MeV}/{\textrm{fm}}^{3}$ as the same as taken in previous works~\cite{Chen:2008PRD,Chen:2011PRD,Chen:2016EPJA,Haymaker:1991RNCSIF}.

The quark matter in a compact star should also be in $\beta$-equilibrium and electric charge neutral,
so we have:
\begin{equation}
\mu_{d}^{} = \mu_{u}^{} + \mu_{e}^{} = \mu_{s}^{} \, ,
\end{equation}
\begin{equation}
\frac{2\rho_{u}^{} - \rho_{d}^{} - \rho_{s}^{}}{3} - \rho_{e}^{} - \rho_{\mu^{-}}^{} = 0 \, .
\end{equation}
And we have the baryon density and chemical potential as:
\begin{equation}
\rho_{B}^{} = \frac{1}{3}(\rho_{u}^{} + \rho_{d}^{} + \rho_{s}^{} ) \, ,
\end{equation}
\begin{equation}\label{muB}
\mu_{B}^{} = \mu_{u}^{} + 2\mu_{d}^{} \, .
\end{equation}

Therefore, we can calculate the properties of the quark matter with a given baryon chemical potential $\mu_{B}$ or baryon density $\rho_{B}$.

In the following sections, we will also use $\rho_{Q}=\rho_{B}$ to denote the baryon density in quark matter.

\section{\label{sec:construction} Construction of the Complete Equation of State and the Speed of Sound}

\subsection{Gibbs Construction}

Since we know little about the detail of hadron-quark phase transition,
we have to consider the hadron phase and the quark phase separately as described above,
and then combine the obtained results together to get the properties of the matter in the phase transition region.

A straight forward way to describe the first-order hadron-quark phase transition is the Maxwell construction.
However, the Maxwell construction can only be applied to a system with only one chemical potential.
Since the neutron star matter consists of two chemical potentials,
	  the baryon chemical potential and electron chemical potential,
	  if we use Maxwell construction, the electron chemical potential will not be continuous at the phase boundary.
For detail, see Ref.~\cite{Glendenning:2000}.

In order to describe the first-order phase transition with two chemical potentials,
   a commonly implemented method for this combination is the Gibbs
construction~\cite{Glendenning:2000,Glendenning:1992PRD,Carroll:2009PRC,Weissenborn:2011ApJL,Fischer:2011ApJS,Schulze:2011PRC,Maruyama:2007PRD}.
In the spirit of Gibbs construction,
there is a mixed region where hadron matter and quark matter coexist.
And although the hadron matter and the quark matter are both charged,
their combination is charge neutral,
i.e. the global charge neutral condition should be guaranteed.

If the volume fraction of quark matter is $\chi$, we have:
\begin{equation}\label{eqn:condition2}
(1-\chi)\rho_{H}^{c} (\mu_{n}^{}, \mu_{e}^{}) + \chi \rho_{Q}^{c}(\mu_{n}^{} , \mu_{e}^{}) = 0 \, ,
\end{equation}
where $\rho_{H}^{c}$ and $\rho_{Q}^{c}$ is the charge density of hadron and quark matter, respectively.

Also, inside the mixed region,
the hadron matter and the quark matter should have the same chemical potential and pressure:
\begin{equation}\label{eqn:condition1}
p_{H}^{}(\mu_{n}^{}, \mu_{e}^{}) = p_{Q}^{} (\mu_{B}^{}, \mu_{e}^{}) \, ,
\end{equation}
where $p_{H}$ and $p_{Q}$ is the pressure for hadron and quark matter, respectively.

Combining Eqs.~(\ref{eqn:condition2}) and (\ref{eqn:condition1}),
together with the equations for hadron and quark matter,
we can solve the baryon chemical potential and electron chemical potential
at a given quark fraction $\chi$.

The energy density $\varepsilon_{M}^{}$ and the baryon number density $\rho_{M}^{}$ for the matter in the phase coexistence region is just the sum of the two phases with the weight of quark fraction:
\begin{equation}
\varepsilon_{M}^{}= \chi \varepsilon_{Q}^{} (\mu_{n}^{},\mu_{e}^{})
 + (1-\chi)\varepsilon_{H}^{} (\mu_{n}^{},\mu_{e}^{}) \, ,
\end{equation}
\begin{equation}
\rho_{M}^{} = \chi \rho_{Q}^{} (\mu_{n}^{}, \mu_{e}^{}) + (1 - \chi) \rho_{H}^{} (\mu_{n}^{} , \mu_{e}^{}) \, ,
\end{equation}
where $\varepsilon_{H}$ and $\varepsilon_{Q}$ are the energy density for hadron and quark matter,
and $\rho_{H}$ and $\rho_{Q}$ are the baryon number density for hadron and quark matter, respectively.

Meanwhile the pressure of the mixed phase equates to that of the hadrons, and so do the quarks, which reads
\begin{equation}
p_{M}^{} = p_{H}^{} = p_{Q}^{} \, .
\end{equation}

\subsection{Sound Speed Construction}

After calculating the EoS with Gibbs construction,
we can calculate the speed of sound by simply adopting $c_{s}^{2}=(\partial p/\partial\varepsilon)$.

\begin{figure}[!htb]
\includegraphics[width=0.50\textwidth]{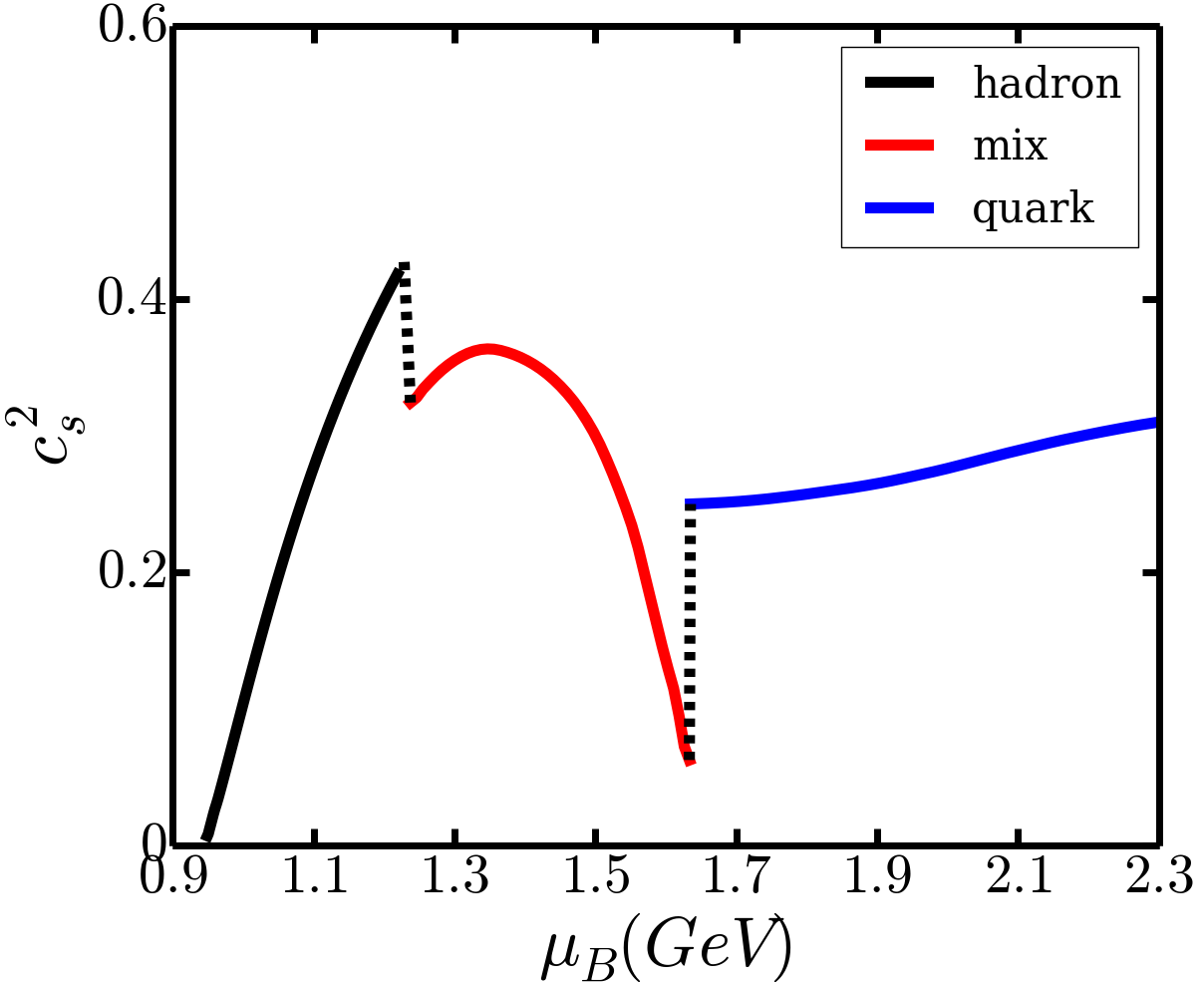}
\vspace*{-10mm}
\caption{Calculated sound speed squared as a function of baryon chemical potential.
The hadron matter is described with the TW99 model in the RMF theory
and contains only protons and neutrons (and leptons).
The quark matter is described with the DS equation of QCD with $\alpha=2$.
The mixed phase (coexistence phase) is described with the Gibbs construction.}
\label{fig:mu-c2_Gibbs}
\end{figure}

The calculated speed of sound as a function of baryon chemical potential is shown in Fig.~\ref{fig:mu-c2_Gibbs}.
Since the matter at low baryon chemical potential is in pure hadron phase, the speed of sound increases monotonically with the increasing of baryon chemical potential.
At $\mu_{B,0}^{}=1.23\textrm{GeV}$, the quark matter begins to appear,
and the speed of sound gets discontinuous.
The corresponding gap of the sound speed squared is $\Delta c_{s}^{2} (\mu_{B,0}^{})=0.103$.
There is another gap in the speed of sound at $\mu_{B,1}=1.63\textrm{ GeV}$,
where the hadron matter totally disappears and the phase transition ends.
The gap of the sound speed squared at such an end is $\Delta c_{s}^{2}(\mu_{B,1}^{})=0.187$.

The Gibbs construction is quite successful,
and we can calculate the baryon chemical potential for the phase transition to take place.
However, as mentioned above, both the hadron model and the quark model are inaccurate at the densities where the phase transition takes place,
so that the above calculated critical baryon chemical potential loses its accuracy.
Also, in our previous work~\cite{Bai:2018PRD}, we have shown that if we take the Gibbs construction to build the complete EoS,
the maximum mass of the hybrid star can not reach $2M_{\odot}$ even if we don't include hyperons.
We have also shown that, if we include hyperons in the hadron model,
the phase transition will not occur under Gibbs construction.
Therefore, we need to find another scheme to construct the EoS of the matter in the phase coexistence region.

In the spirit of 3-window construction,
we should interpolate the hadron model and the quark model to get the EoS of matter in the middle density region.
There have developed methods to interpolate the $p_{H}^{}(\mu_{B})$ and $p_{Q}^{}(\mu_B)$ using a polynomial function~\cite{Kojo:2015PRD},
and to interpolate the $\varepsilon_{H}(\rho_{B}^{})$ and $\varepsilon_{Q}(\rho_{B}^{})$ using hyperbolic weight function~\cite{Masuda:2013PTEP}.
However, these interpolation scheme cannot take into consideration of the characteristics of the
first-order phase transition.

Therefore, in this work, we propose to make use of the variation feature of the sound speed
to construct the equation of state,
and we will calculate various quantities in the corresponding argument space to look over
how astronomical observations can constrain the interpolation parameters.

There have been different models to construct or calculate the speed of sound in dense matter
(see, e.g. Refs.~\cite{Alford:2013PRD,Zheng:2015APJ,Zhang:2016PRC,Tews:2018APJ,Han:2018arXiv,Aloy:2019MNRAS,Annala:2019arXiv}).
In this paper, the speed of sound is calculated using the RMF model in low chemical potential region,
and using the DS equation approach in high chemical potential region.
And in the middle region, inspired by the result via Gibbs construction,
the speed of sound is constructed with a quadratic function:
\begin{equation}
c_{M}^{2}(\mu_B)=A\mu_{B}^{2}+B\mu_{B}+C,
\end{equation}
where $A$, $B$ and $C$ are parameters describing the sound speed in the phase coexistence region.
For simplicity of representation, we denote the phase transition (coexistence) region as $\mu_{0}^{}\le\mu_{B}\le\mu_{1}^{}$,
where $\mu_{0}^{}$ and $\mu_{1}^{}$ correspond to the beginning and the ending of the phase transition.

After constructing the speed of sound as a function of baryon chemical potential
with five parameters $A$, $B$, $C$, $\mu_{0}$ and $\mu_{1}$,
we can calculate the equation of state by solving the equations
\begin{eqnarray}
\frac{\partial \rho_{B}^{}}{\partial \mu_{B}^{}} \! \! & \! \! = \! \! & \! \! \frac{\rho_{B}^{}}{\mu_{B}^{}c_{s}^{2}(\mu_{B}^{})},\\
\frac{\partial p}{\partial \mu_{B}^{}} \! \! & \! \! = \! \! & \! \! \rho_{B}^{} \, ,
\end{eqnarray}
which are just the simple thermodynamic relations.
The boundary condition is
\begin{eqnarray}
\rho_{B}^{}(\mu_{0}) \! \! & \!\! = \rho_{H}^{}(\mu_{0}),\\
p(\mu_{0}) \!\! & \!\! = p_{H}^{}(\mu_{0}) \, ,
\end{eqnarray}
where $\rho_{H}$ and $p_{H}$ is the baryon number density and the pressure calculated using hadron model.

However, in high density region, although the sound speed is the same,
the constructed EoS and quark EoS may have a constant difference.
Therefore, we require:
\begin{eqnarray}\label{eqn:boundary2}
		p(\mu_{1})=p_{Q}^{}(\mu_{1})\,,\\
		\rho_{B}(\mu_{1})= \rho_{Q}^{}(\mu_{1})\,,\label{eqn:boundary2-2}
\end{eqnarray}
where $p_{Q}^{}$ and $\rho_{Q}^{}$ is the pressure and the baryon number density of quark matter, respectively.
We take Eqs.~(\ref{eqn:boundary2}) and (\ref{eqn:boundary2-2}) to constrain the parameters,
and then only three of the five parameters are independent.

\section{\label{sec:numerical}Numerical Results and Discussions}
\subsection{Constructed EoS and Parameter Space}

In order to fix the interpolation of the sound speed squared,
we need three of the five parameters: $A$, $B$, $C$, $\mu_{0}$ and $\mu_{1}$, because of the quadratic relation,
or the $\mu_{0}$, $\mu_{1}$, $\Delta c_{s}^{2}(\mu_{0})$, $\Delta c_{s}^{2}(\mu_{1})$,
and the $c_{s}^{2}$ at a chemical potential $\mu \in (\mu_{0} , \mu_{1} )$.
In this work, we take the sound speed squared $c_{s}^{2}(\mu_{\scriptsize\textrm{middle}})$ to be a free parameter, where $\mu_{\scriptsize \textrm{middle}}=(\mu_{0}+\mu_{1})/2$.
As analyzed above, only three of these five parameters are independent.
We take them as $\mu_{0}$, $\mu_{1}$ and $c_{s}^{2}(\mu_{\scriptsize\textrm{middle}})$.
Some of the parameter set may result in the same quadratic function,
but the phase transition region would be different.

In our calculation, the ranges of these parameters are set as
$0.95\le \mu_{0} \le 1.5 \textrm{GeV}$,
$1.2 \textrm{GeV}\le \mu_{1} \le 2.0 \textrm{GeV}$,
$0 \le c_{\scriptsize \textrm{middle}}^{2} < 1.0$.
We randomly choose these parameters from their corresponding ranges,
and construct the square of the sound speed, and deduce further the EoS in each case.
We do this for 200000 times, and get a large number of values of the sound speed squared and the EoS.
%
%All the possible speed of sound and EoS are shown in Fig.~\ref{fig:mu_cs2_p_e_freq_Nq_ALL}.
%
%Notice that the parameters which will result in  $c_{s}^{2}>1$ or $c_{s}^{2}<0$ have already been excluded.
%

It is known that, at the starting chemical potential of the phase transition,
there is a gap in the speed of sound, and its value can be written as:
\begin{equation}
	\Delta c_{0}^{2} = c_{H}^{2}(\mu_{0})-c_{M}^{2}(\mu_{0}),
\end{equation}
where $c_{H}^{2}(\mu)$ is the sound speed squared calculated through the hadron model,
$c_{M}^{2}(\mu)$ is the one constrained for the matter in the phase coexistence region.

Similarly, at the ending of the phase transition, we have
\begin{equation}
	\Delta c_{1}^{2}=c_{Q}^{2}(\mu_{1})-c_{M}^{2}(\mu_{1}),
\end{equation}
where $c_{Q}^{2}(\mu)$ is the sound speed squared calculated with the DS equation approach.
Since the constructed sound speed should be qualitatively the same as that calculated
with the Gibbs construction,
we require then $\Delta c_{0}^{2}>0$ and $\Delta c_{1}^{2}>0$ (see Fig.~\ref{fig:mu-c2_Gibbs}).

\begin{figure*}[!htp]
\vspace*{-2mm}
\includegraphics[width=1.0\textwidth]{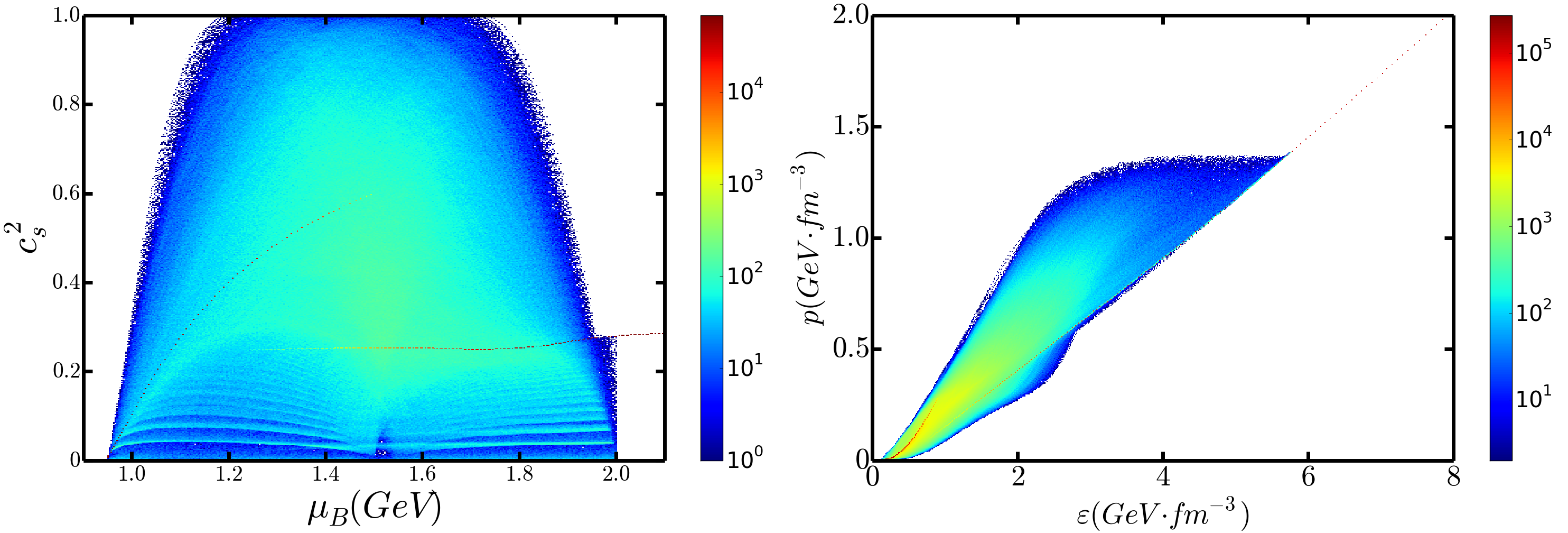}
\vspace*{-6mm}
\caption{{\it Left panel:} The 2-dimensional histogram for all the possible $c_{s}^{2}(\mu_{B})$ function constructed using our interpolation scheme.
The hadron matter is described with the TW-99 model of the RMF theory without the inclusion of hyperons.
The quark matter is described with the DS equation approach of QCD with $\alpha=2$.
The different color corresponds to the number of different $c_{s}^{2}(\mu_{B})$ functions passing through a certain point.
{\it Right panel:} The 2-dimensional histogram for all the correspondingly constructed EoSs.
%
%The hadron and quark matter are the same as in left panel.
%
	The color bar corresponds to the number of different EoSs passing through a certain point.
	}
\label{fig:mu_cs2_p_e_freq_Nq_ALL}
\end{figure*}

In Fig.~\ref{fig:mu_cs2_p_e_freq_Nq_ALL}, we show all the constructed functions $c_{s}^{2}(\mu_{B})$ and the EoSs.
The different color corresponds to the number of the $c_{s}^{2}(\mu_{B})$ functions or the EoSs passing through a certain point.
As we can see from the figure, the sound speed squared in the phase transition region varies from 0 to 1,
which means we have included both the very soft and the very stiff EoSs.

After having the EoS of the dense matter,
we can calculate the mass-radius relation of the compact star composed of the matter
by solving the TOV equation
\begin{eqnarray}
		\frac{\textrm{d}p}{\textrm{d}r} \! \! & \! \! = \! \! & \!\! -\frac{m\varepsilon}{r^{2}}\frac{(1+p/\varepsilon)(1+4\pi r^{3}p/m)}{1-m/r},\\
		\frac{\textrm{d}m}{\textrm{d}r} \! \! & \! \! = \! \! & \! \! 4\pi r^{2}\varepsilon,
\end{eqnarray}
where we have taken the $G=1$ unit.
%
%This equation can be solved with a given central pressure,
%and integrating from the center to the surface of the compact star, where the pressure vanishes.
%

The tidal deformability $\Lambda$ can be related to the mass $M$ and the radius $R$ of the star as
\begin{equation}
\Lambda = \frac{2}{3}k_{2}\left(\frac{R}{M}\right)^{5}\,,
\end{equation}
where $k_{2}$ is the tidal love number of the compact star, which can be calculated
as~\cite{Flanagan:2008PRD,Hinderer:2008ApJ,Postnikov:2010PRD}
\begin{equation}
		k_{2}=\frac{8}{5}\beta^{5}z[6\beta(2-y_{R})+6\beta^{2}(5y_{R}-8)+4\beta^{3}(13-11y_{R})
		+4\beta^{4}(3y_{R}-2)+8\beta^{5}(1+y_{R})+3z\textrm{ln}(1-2\beta)]^{-1}\,,
\end{equation}
where $\beta=M/R$ is the compactness parameter,
\begin{equation}
	z\equiv(1-2\beta^{2})(2-y_{R}+2\beta(y_{R}-1)),
\end{equation}
and $y_{R}=y(r=R)$ and $y(r)$ is calculated by solving the equation
\begin{equation}
	\frac{\textrm{d}y}{\textrm{d}r}=-\frac{y^{2}}{r}-\frac{y-6}{r-2m}-rQ,
\end{equation}
where
\begin{equation}
		Q \equiv  \ 4\pi\frac{(5-y)\varepsilon+(9+y)p+(\varepsilon+p)/c_{s}^{2}}{1-2m/r}
		  -\left[\frac{2(m+4\pi r^{3}p)}{r(r-2m)}\right]^{2}.
\end{equation}

\begin{figure*}[!htb]
\vspace*{-3mm}
\center
\includegraphics[width=0.51\textwidth]{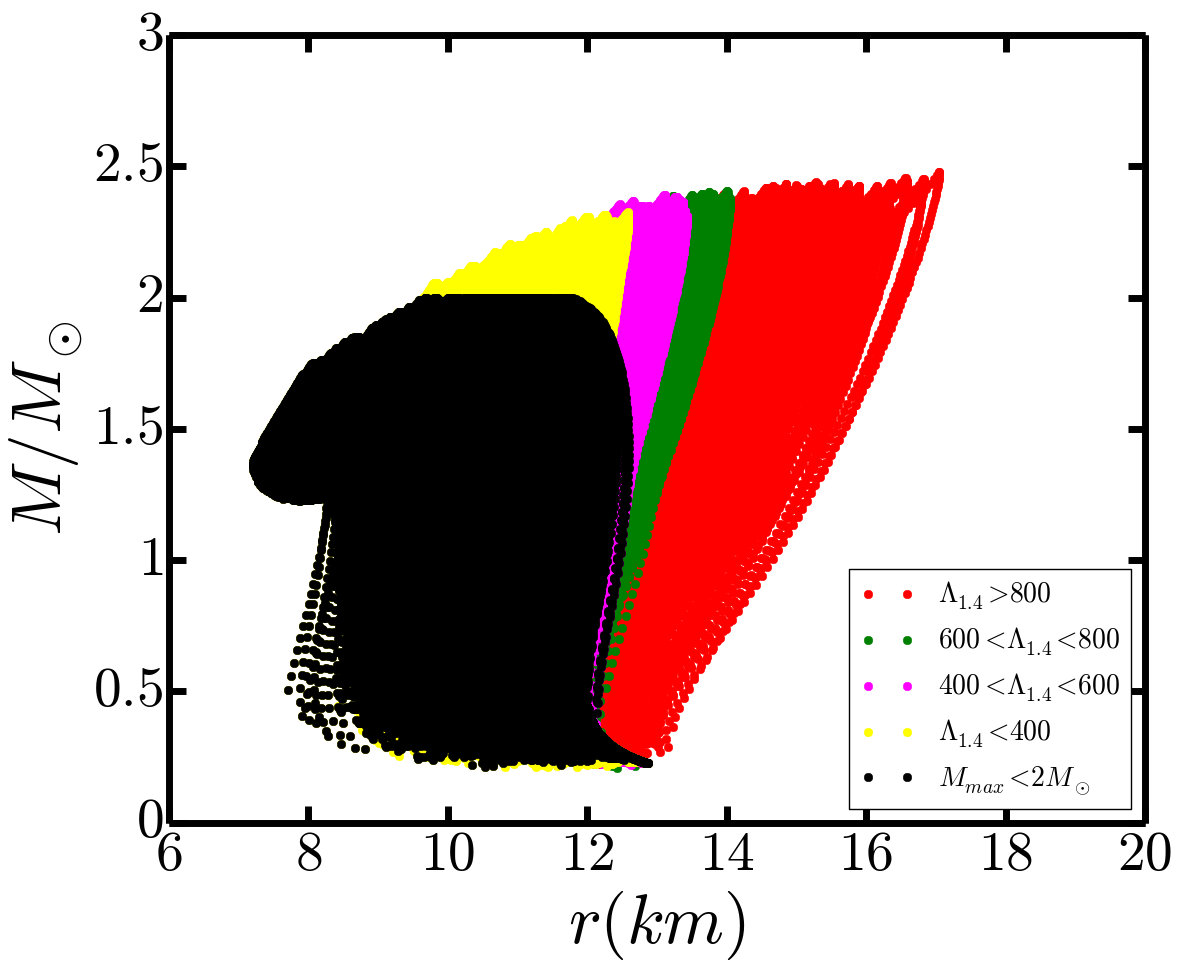}
\vspace*{-3mm}
\caption{(color online) Calculated mass-radius relation of the neutron stars from various interpolated EoSs.
In all the EoSs, the hadron matter is described using the TW-99 model without hyperons,
and quark matter is described via the DS equation approach with $\alpha=2$.}
\label{fig:MassRadiusCloud}
\end{figure*}

The calculated mass-radius relation with all the constructed EoSs is shown in Fig.~\ref{fig:MassRadiusCloud}.
The maximum mass of our construction can be $M_{\textrm{max}}\le 2.48\,M_{\odot}$.
For a $1.4M_{\odot}$ hybrid star,
the range of the radius and the tidal deformability are:
\begin{eqnarray}
		8.18\le \!\! & \! \! R_{1.4} \le 15.50 \, \textrm{km}\,,\\
		14.83\le \!\!\! &\! \! \Lambda_{1.4}\le 2558.11. \;
\end{eqnarray}
%
%Notice that this is not our prediction on the range of mass, radius and tidal deformability.
%On the contrary, we are going to use different ranges of these quantities to constrain our parameter.
%

It is clear from the figure that if we require that the tidal deformability to be smaller,
the radius of the hybrid star will also be smaller,
and so does the maximum mass of the hybrid star correspondingly.
Therefore, a small tidal deformability and a large neutron star mass together provide strong constraints on the EoS.

\subsection{Astronomical Constraints on the Parameters}

We have constructed a lot of different EoSs.
However, not every EoS satisfies the constraints of astronomical observations.
Since neutron star with a mass of about $2M_{\odot}$ has already been observed~\cite{Demorest:2010Nature,Antoniadis:2013Science},
we should exclude those constructed EoS which result in a maximum mass less than $2M_{\odot}$.
The detection of gravitational wave GW170817 also sets the upper limit of neutron star maximum mass,
so we require that the maximum mass is less than $2.16M_{\odot}$~\cite{Margalit:2017ApJ,Rezzolla:2018ApJ,Ruiz:2018PRD}.
The gravitational wave has provided information on the tidal deformability $\Lambda$.
For a $1.4M_{\odot}$ neutron star, we require then $\Lambda_{1.4}<800$ according to Ref.~\cite{Abbott:2017PRL},
and $\Lambda_{1.4}>120$ according to Ref.~\cite{Annala:2018PRL},
where a wide range of different EoSs is considered.
Since the neutron star radius depends on the model,
we take $9.9<R<13.6\textrm{km}$ for $1.4M_{\odot}$ neutron star according to Ref.~\cite{Annala:2018PRL}
to constrain the EoS.

\begin{figure*}[!htp]
\includegraphics[width=0.98\textwidth]{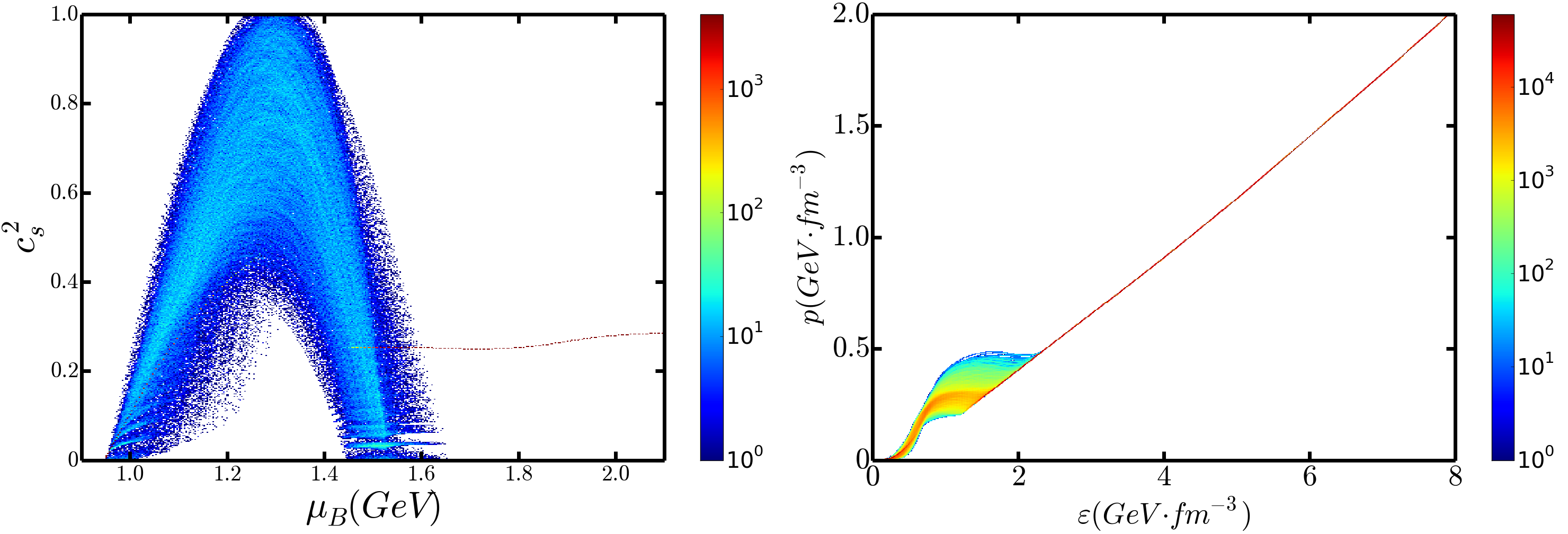}
\vspace*{-3mm}
\caption{{\it Left panel:} The histogram for $c_{s}^{2}(\mu_{B})$ functions satisfying the condition: $2<M_{\textrm{max}}/M_{\odot}<2.16$, $120<\Lambda_{1.4}<800$ and $9.9<R_{1.4}<13.6\textrm{km}$.
	The hadron matter is described with the TW-99 model without the inclusion of hyperons.
	The quark matter is described by the DS equation approach with $\alpha=2$.
	The different color corresponds to the number of the $c_{s}^{2}(\mu_{B})$ functions
    passing through a certain point.
	{\it Right panel:} The same as the left panel but for the EoSs.
% satisfying the same astronomical constraints as the left panel.
%	The theoretical models describing the different matters are also, respectively, the same as in left panel.
	}
\label{fig:mu_cs2_p_e_freq_Nq_constrain}
\end{figure*}

After the calculated maximum mass of neutron star and the tidal deformability and radius of the star
with a mass $1.4M_{\odot}$ being constrained with the above mentioned observations,
only a small number of the constructed sound speed squared and the EoS survive.
The obtained results are shown in Fig.~\ref{fig:mu_cs2_p_e_freq_Nq_constrain}.

In the left panel of Fig.~\ref{fig:mu_cs2_p_e_freq_Nq_constrain},
we show the constructed sound speed squared satisfying the astronomical observations.
It can be seen from the figure that the baryon chemical potential for the hadron-quark phase transition to occur can be constrained by the astronomical observations.
In our interpolation scheme, we have parameters $\mu_{0}^{}$ and $\mu_{1}^{}$ which correspond to the beginning, the ending of the phase transition, respectively.
We analyze then the $\mu_{0}^{}$ and the $\mu_{1}^{}$ dependence of the number of EoSs satisfying
the astronomical constraints in the following.

\begin{figure*}[!htb]
\vspace*{-1mm}
\center
\includegraphics[width=0.53\textwidth]{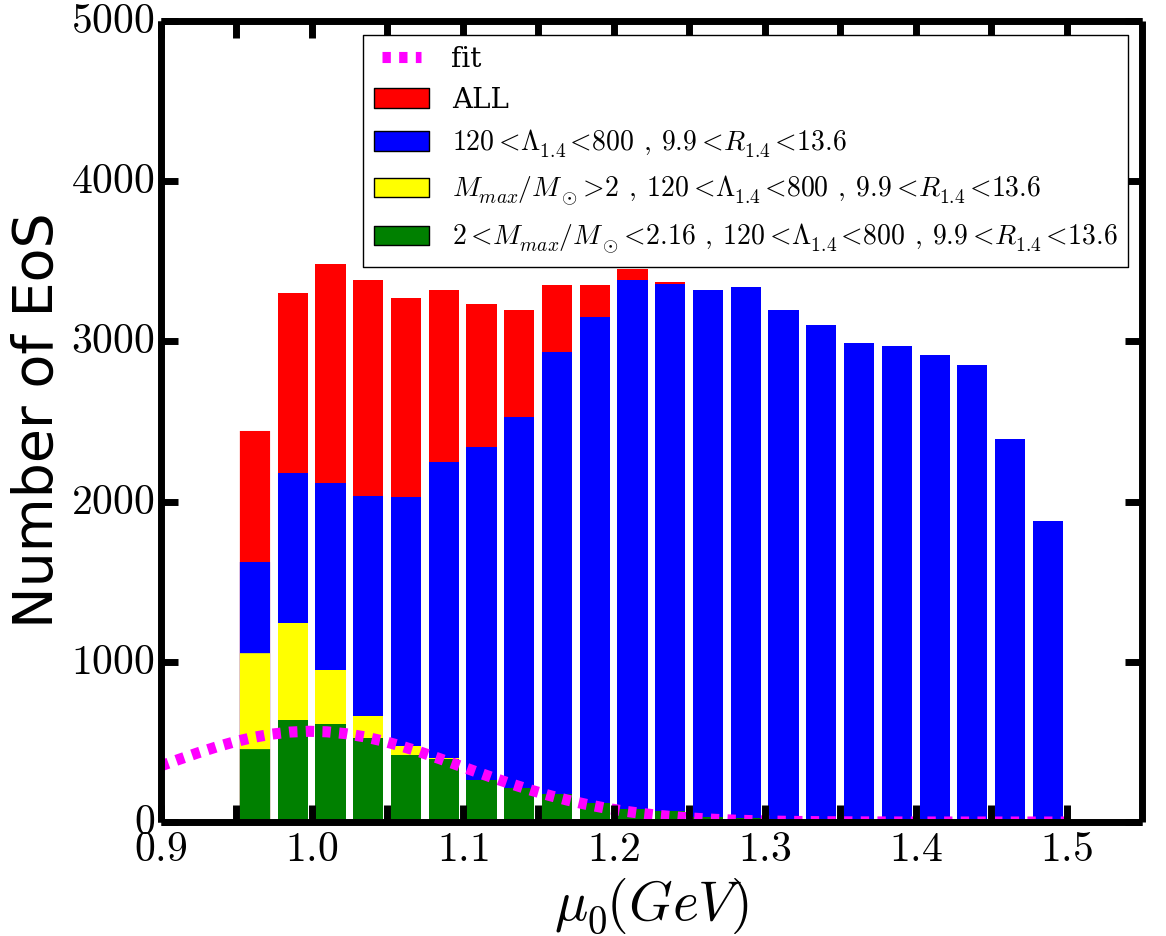}
\vspace*{-3mm}
\caption{Calculated $\mu_{0}$ dependence of the number of constructed EoSs constrained with several kinds of astronomical observations. In the construction, the hadron matter is described with the TW-99 model without hyperons,
and the quark matter is described via the DS equation approach with $\alpha=2$.
Different colors corresponds to different astronomical constraints.
The pink dotted line is the fit of the green bars using Gaussian distribution.}
\label{fig:mu0_freq_Nq}
\end{figure*}

In Fig.~\ref{fig:mu0_freq_Nq}, we show the $\mu_{0}^{}$ dependence of the number of EoSs without any astronomical constraints with red bars,
those satisfying the requirement $120<\Lambda_{1.4}<800$, $9.9<R_{1.4}<13.6\textrm{km}$
with blue bars,
those with further constraint $M_{\scriptsize \textrm{max}} > 2 M_{\odot}$ with yellow bars,
and the ones with much further requirement $ 2 M_{\odot} < M_{\scriptsize \textrm{max}} < 2.16M_{\odot}$ in green bars.
From Fig.~\ref{fig:mu0_freq_Nq}, one can find easily that, although the constraints on the tidal deformability
and the radius reduce the number of the EoSs for different values of the $\mu_{0}^{}$,
they do not change the range of the $\mu_{0}^{}$.
Meanwhile the lower limit of the maximum mass reduces the upper limit of the $\mu_{0}^{}$,
and the upper limit of maximum mass doesn't change the range of $\mu_{0}^{}$ either.
After taking into account all the astronomical constraints,
the range of $\mu_{0}^{}$ is constrained to be $\mu_{0}^{}\le 1.34\,\textrm{GeV}$,
which corresponds to baryon number density $n_{0}\le 4.26n_{s}$,
where $n_{s}$ is the saturation density of the nuclear matter.
However, the lower limit of $\mu_{0}^{}$ is set as $0.95\,\textrm{GeV}$,
which corresponds to nearly zero baryon number density.
This means that, to constrain the lower limit of $\mu_{0}^{}$ better with the astronomical observations,
we should set the initial value of such a limit smaller.

Since our initial parameters are randomly distributed in their range,
the number of the EoSs should be proportional to the probability distribution.
Therefore, we make use of a Gaussian distribution function to fit the number distribution of the green bars
in Fig.~\ref{fig:mu0_freq_Nq},
and the fitted result is plotted in the figure with pink dotted line.
The most probable value of the $\mu_{0}^{}$ is $\langle\mu_{0}^{}\rangle =1.00\,\textrm{GeV}$,
corresponding to a baryon number density $1.5n_{s}$ at which the nucleons in the matter begin to overlap with each other~\cite{Liu:2001NPA}.

\begin{figure*}[!htb]
\vspace*{-2mm}	
\center
\includegraphics[width=0.51\textwidth]{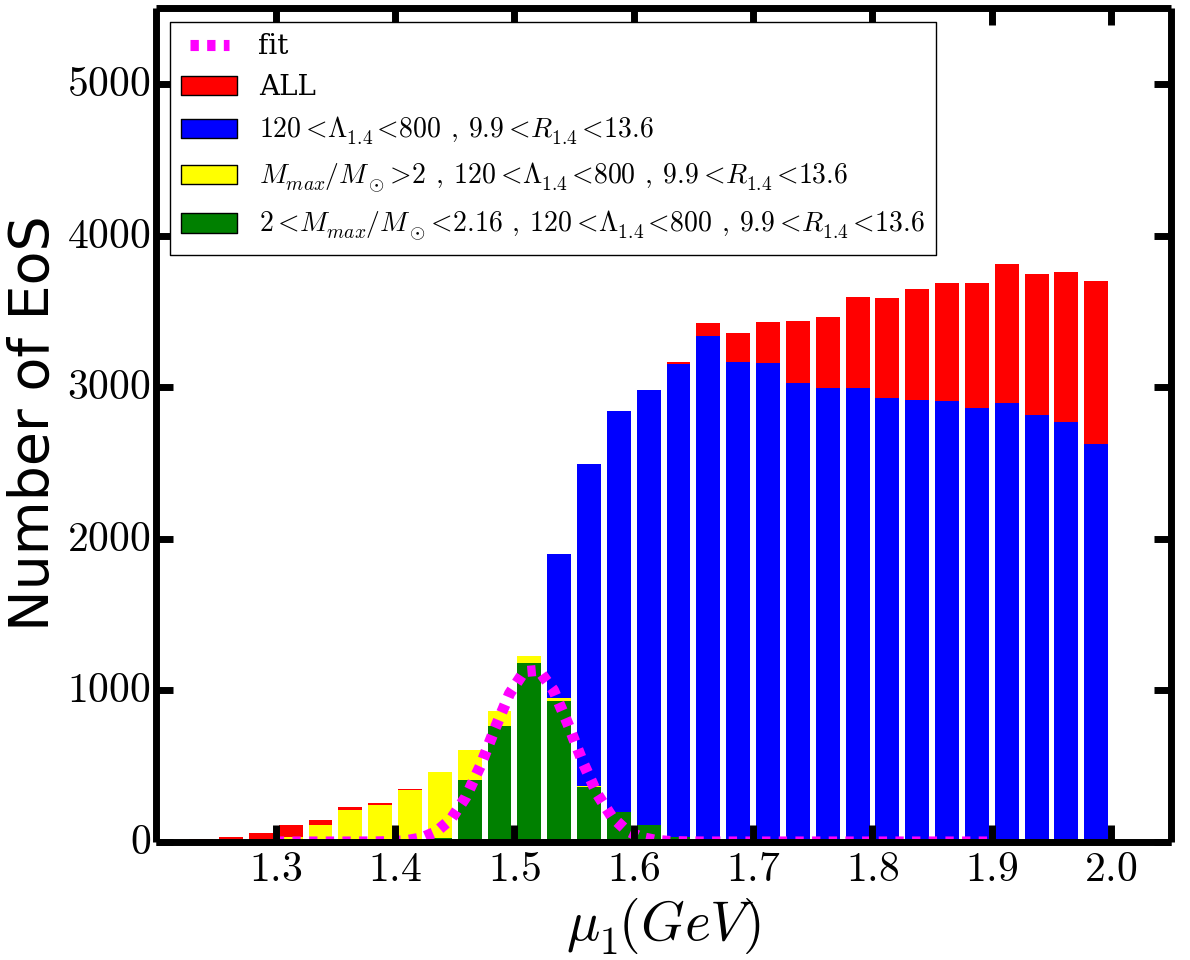}
\vspace*{-3mm}	
\caption{The same as Fig.~\ref{fig:mu0_freq_Nq} but for the $\mu_{1}$ dependence.
	}
\label{fig:mu1_freq_Nq}
\end{figure*}

The similar analysis can also be carried out on the $\mu_{1}^{}$,
the baryon chemical potential corresponding to the ending of the hadron-quark phase transition.
The obtained results of the $\mu_{1}^{}$ dependence of the number of the EoSs satisfying different astronomical constraints are displayed in Fig.~\ref{fig:mu1_freq_Nq} with the same notations as those in
Fig.~\ref{fig:mu0_freq_Nq}.

Although the parameters are taken randomly in their range,
the red bars which label the number of the EoSs without any astronomical constraint shown
in Fig.~\ref{fig:mu1_freq_Nq} is not uniformly distributed.
This is because when $\mu_{1}^{}$ is too small,
there are possibilities for $\mu_{1}^{} < \mu_{0}^{}$,
or the construction cannot satisfy the boundary condition in Eqs.~(\ref{eqn:boundary2}) and (\ref{eqn:boundary2-2}).
The lower limit of the $\mu_{1}^{}$ with the constraints on the tidal deformability and the radius
is larger than that without the constraints
(notice that the blue bars in low $\mu_{1}^{}$ region are completely hidden behind the yellow bars).
The requirement that $M_{\scriptsize \textrm{max}}>2M_{\odot}$ reduces the upper limit of $\mu_{1}^{}$ (yellow bars),
while the upper limit of maximum mass increases the lower limit of $\mu_{1}^{}$ (green bars).
Therefore, after applying the astronomical constraints,
the range of $\mu_{1}^{}$ is assigned as $1.44\le \mu_{1}^{}\le 1.65\,\textrm{GeV}$,
corresponding to a baryon number density range $6.41\le n_{1}/n_{s}\le 11.26$.
As the same as done in Fig.~\ref{fig:mu0_freq_Nq}, we fit the green bars with Guassian distribution function,
and find that the most probable chemical potential is $\langle\mu_{1}^{}\rangle=1.51\,\textrm{GeV}$,
which correspond to $\langle n_{1}\rangle =7.90n_{s}$.

\subsection{Inclusion of Hyperons and other astronomical constraints}

\begin{figure*}[!htb]
\vspace*{-1mm}
\center
\includegraphics[width=0.51\textwidth]{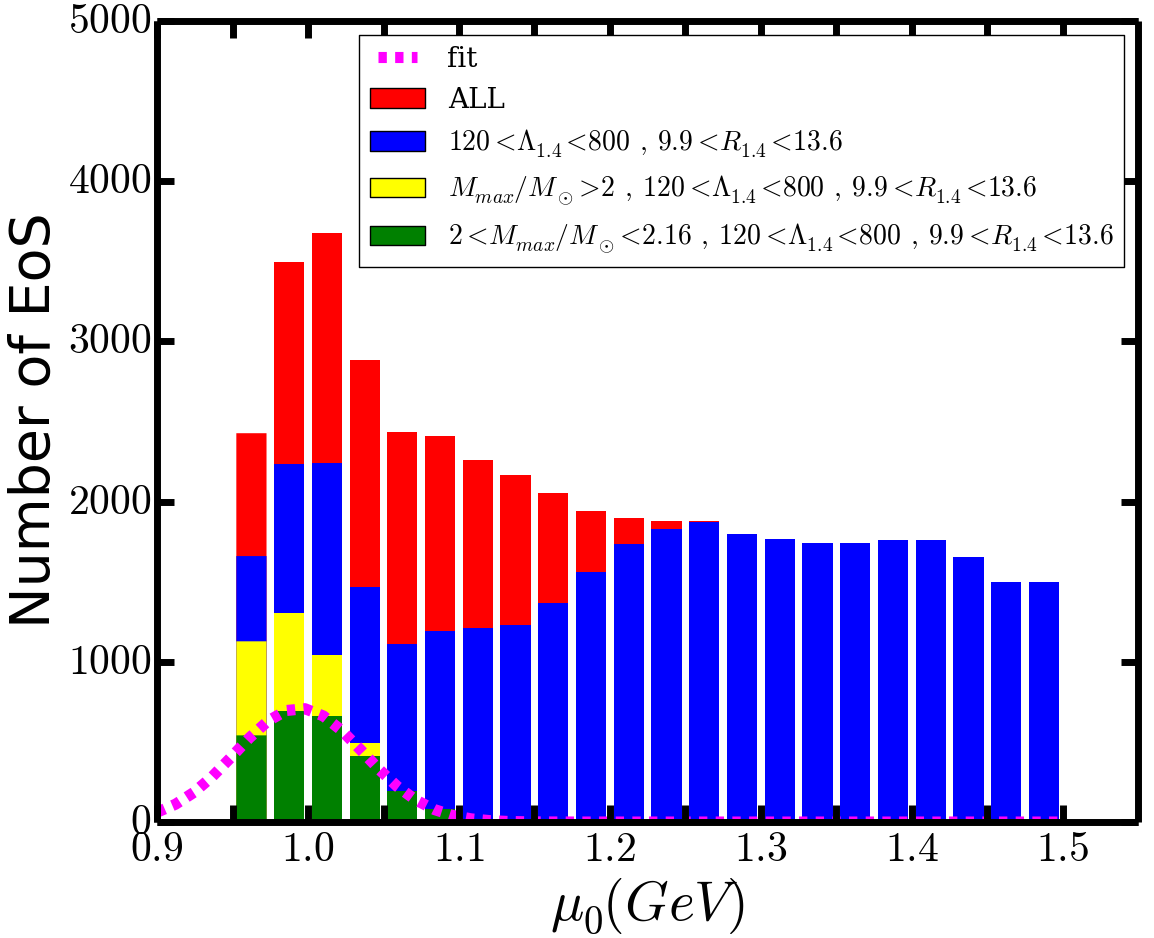}
\vspace*{-3mm}
\caption{The same as Fig.~\ref{fig:mu0_freq_Nq} but for the case that hyperons are included in the hadron model.
	}
\label{fig:mu0_freq_NYq}
\end{figure*}

\begin{figure*}[!htb]
\vspace*{-1mm}
\center
\includegraphics[width=0.51\textwidth]{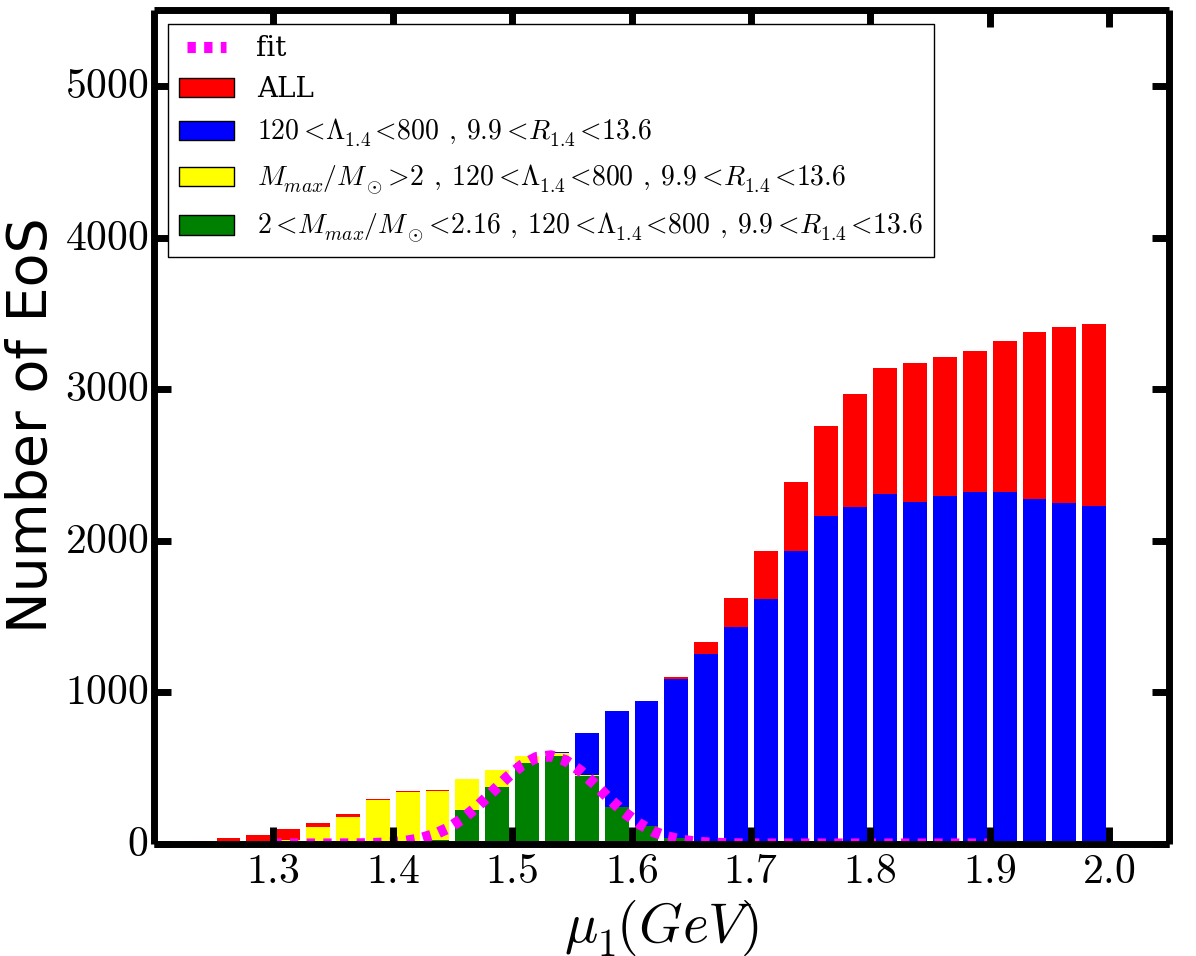}
\vspace*{-3mm}
\caption{The same as Fig.~\ref{fig:mu1_freq_Nq} but for the case that hyperons are included in the hadron model.
	}
\label{fig:mu1_freq_NYq}
\end{figure*}

We have already shown that astronomical observations can help constrain the chemical potentials labelling the phase transition region.
However, for hadron matter, we have only included protons and neutrons for the baryons in the above discussion.
We are going to look through the effects of hyperons in the following.
In Figs.~\ref{fig:mu0_freq_NYq} and \ref{fig:mu1_freq_NYq},
we show the $\mu_{0}$, $\mu_{1}$ dependence of the number of EoSs under the astronomical constraints
when hyperons are included, respectively.

%\begin{widepage}

\begin{center}
\begin{table*}[!htb]
\caption{Constrained quantities featuring the hadron-quark phase transition under different astronomical observations. The composition Nq, NYq refers to the case that the hadron matter sector does not include,
or include hyperons, respectively.
The baryon chemical potentials are in units GeV, and the baryon number densities are in unit $n_{s}$.}
\label{tab:other_constraints}
	\resizebox{\textwidth}{!}
	{
\begin{tabular}{cccccccccc}
%\hline
%\multicolumn{3}{c}{Astronomical observations} & &\multicolumn{10}{c}{Constrained Range of the quantities}\\
\hline
$M_{\textrm{max}}(M_{\odot})$ & $\Lambda_{1.4}$&$R_{1.4}(\textrm{km})$ & composition & $\mu_{0\textrm{max}}$ & $n_{0\textrm{max}}$    &$\mu_{1\textrm{min}}$&$\mu_{1\textrm{max}}$&$n_{1\textrm{min}}$&$n_{1\textrm{max}}$\\
\hline
	\multirow{2}{*}{2-2.16}& \multirow{2}{*}{120-800}&\multirow{2}{*}{9.9-13.6}                         &Nq &1.34&4.26&  1.44&1.66&6.41&11.27 \\
						   &                         &                                           &NYq&1.12&3.15&  1.42&1.67&6.07&11.64 \\
\hline
	\multirow{2}{*}{2-2.35}& \multirow{2}{*}{120-800}&\multirow{2}{*}{9.9-13.6}                         &Nq &1.34&4.26&  1.31&1.66&4.58&11.27  \\
						   &                         &                                           &NYq&1.12&3.15&  1.31&1.67&4.57&11.64  \\
\hline
	\multirow{2}{*}{2-2.16}& \multirow{2}{*}{344~\cite{Malik:2018PRC}-800}&\multirow{2}{*}{9.9-13.6}    &Nq &1.34&4.26&  1.44&1.56&6.41&8.93   \\
						   &                         &                                          &NYq&1.04&2.15&  1.42&1.57&6.07&9.05    \\
\hline
\multirow{2}{*}{2-2.16}& \multirow{2}{*}{120-800}&\multirow{2}{*}{10.62-12.83~\cite{Lattimer:2014EPJA}} &Nq &1.34&4.26&  1.44&1.62&6.41&10.38   \\
						   &                         &                                          &NYq&1.12&3.15&  1.43&1.62&6.28&10.47     \\
\hline
	\multirow{2}{*}{2-2.16}& \multirow{2}{*}{344-580~\cite{Abbott:2018PRL}}&\multirow{2}{*}{10.62-12.83}&Nq &1.34&4.26&  1.44&1.56&6.49&8.93 \\
						   &                         &                                         &NYq&1.04&2.15&  1.44&1.57&6.48&9.05 \\
\hline
\end{tabular}
	}
\end{table*}
\end{center}

\vspace*{-6mm}

From Fig.~\ref{fig:mu0_freq_NYq}, one can notice that the number distribution of $\mu_{0}$
can still be constrained by the astronomical observations when hyperons are included in the hadron matter.
After considering the requirements $2<M_{\scriptsize \textrm{max}}/M_{\odot}<2.16$, $120<\Lambda_{1.4}<800$ and $9.9<R_{1.4}<13.6$,
the upper limit of the $\mu_{0}$ can be constrained as $\mu_{0}\le 1.12\,\textrm{GeV}$,
which corresponds to baryon number density $n_{0}\le 3.15\,n_{s}$.
%
%However, similar to the case without including hyperons,
%the lower limit of the $\mu_{0}$ can not be determined by the astronomical observations either.
%
After fitting the green bars with Gaussian distribution function,
we find that the most probable chemical potential for the phase transition to begin
is $\langle\mu_{0}\rangle = 0.99\, \textrm{GeV}$,
corresponding to $\langle n_{0}\rangle =1.43\, n_{s}$.
Comparing with the case without hyperons,
one can find that the inclusion of hyperons reduces the upper limit of $\mu_{0}$ obviously
and the most probable value of the $\mu_{0}$ slightly.

Fig.~\ref{fig:mu1_freq_NYq} shows apparently that, after taking the requirement  $2<M_{\scriptsize \textrm{max}}/M_{\odot}<2.16$, $120<\Lambda_{1.4}<800$ and $9.9<R_{1.4}<13.6$,
the range of the $\mu_{1}$ can be constrained as $1.42\le \mu_{1}\le 1.67\, \textrm{GeV}$,
which corresponds to $6.07<n_{1}/n_{s}<11.64$.
By fitting the green bars with Gaussian distribution function,
we find that the most probable $\mu_{1}$ is $\langle\mu_{1}\rangle=1.53\, \textrm{GeV}$,
corresponding to $\langle n_{1}\rangle=8.20\, n_{s}$, which is quite close to that given
in other QCD model calculation~\cite{Chang:2005NPA}.
Comparing with the case without hyperons,
one can observe that the range of the $\mu_{1}$ is slightly extended after including hyperons.
%
%but the change is not big.

We have shown that the astronomical observations can help constrain the baryon chemical potential for the hadron-quark phase transition to occur.
However, the ranges of the astronomical observables, e.g., the mass, the tidal deformability and the radius
have not yet been fixed concretely, and different studies give distinct results.
Therefore, we repeat the above described analyzing process with different astronomical constraints.
The obtained results of the baryon chemical potential and the corresponding baryon density for the hadron-quark phase transition to happen are listed in Table~\ref{tab:other_constraints}.

The first set of Table~\ref{tab:other_constraints} gives the result we have just obtained.
The second set lists the result when a larger upper limit of the maximum mass of a star is taken.
It is evident that, the lower limit of the $\mu_{1}$ reduces a lot, no matter whether hyperons are included.
The third set shows the result under a larger lower limit of the tidal deformability.
It manifests clearly that the upper limits of the $\mu_{1}$ is reduced.
Meanwhile the upper limit of the $\mu_{0}$ is also reduced when hyperons are included.

It has been known that the neutron star radius may be restricted to a narrower range $(10.62\,\textrm{--}\, 12.83)\,$km~\cite{Lattimer:2014EPJA}.
The obtained result under such a further constraint is shown in the fourth set.
One can notice from the data that, the range of the $\mu_{0}$ and the $\mu_{1}$
do not change much correspondingly.
This means that the constraints from the mass and the tidal deformability are more stringent.

%Recent analysis has also provided a narrower range for the tidal deformability\cite{Abbott:2018PRL}.
Recently, in Ref.~\cite{Abbott:2018PRL}, the tidal deformability is estimated to be $70\le\Lambda_{1.4}\le 580$.
The upper limit of this approximation is smaller than what we have used,
while the lower limit is also smaller.
However, since we intend to study the effect of a narrower tidal deformability range,
we shall use a larger lower limit of $\Lambda_{1.4}$ as given in Ref.~\cite{Malik:2018PRC}.
The obtained result under the most strict astronomical constraints is given in the last set of Table~\ref{tab:other_constraints}.
It is apparent that the ranges of the $\mu_{0}$ and the $\mu_{1}$ are narrowed down correspondingly.
It indicates that the chemical potential corresponding to the ending of the hadron-quark phase transition is not influenced much by the inclusion of hyperons under such an astronomical circumstance,
but the chemical potential corresponding to the beginning of the phase transition is obviously reduced by hyperons.

\section{\label{sec:sum}Summary and Remarks}

In this work, we developed a novel scheme to interpolate the hadron and the quark models to construct
a complete EoS for the compact star matter involving hadron-quark phase transition.
We take the speed of sound in the matter as the objective quantity to be interpolated,
with which one can take into consideration the characteristics of the first-order phase transition,
since the sound speed changes abruptly at both the beginning and the ending points of the phase transition.
To describe the hadron matter we take the RMF model with the TW-99 parameters,
and for the quark matter we implement the Dyson-Schwinger equation approach of QCD.

There are five parameters in our interpolation scheme.
They are the baryon chemical potentials at the beginning and the ending of the phase transition
($\mu_{0}$, $\mu_{1}$, respectively),
the gap of the sound speed squared $c_{s}^{2}$ at the $\mu_{0}$ and that at the $\mu_{1}$,
and the $c_{s}^{2}$ at the middle point between the $\mu_{0}$ and the $\mu_{1}$.
With a quadratic relation is proposed for the sound speed in terms of the chemical potential,
the number of the independent parameters is reduced to three.
With the astronomical observation constraints such as the maximum mass of a neutron star,
the radius and the tidal deformability of the star with 1.4 times the solar mass being taken as calibrations,
the baryon chemical potentials which correspond to the beginning and the ending of the phase transition
are determined.

As we take $2<M_{\textrm{max}}/M_{\odot}<2.16$, $120<\Lambda_{1.4}<800$ and $9.9<R_{1.4}<13.6\textrm{km}$
as the constraints, the phase transition chemical potentials of the star matter whose hadron sector does not involve hyperons is assigned as $\mu_{0}\le 1.34\,\textrm{GeV}$, $1.44<\mu_{1}<1.66\,\textrm{GeV}$.
And those for the star matter whose hadron sector involves hyperons are
$\mu_{0}<1.12\,\textrm{GeV}$, $1.43<\mu_{1}<1.67\,\textrm{GeV}$.
The corresponding phase transition baryon number density is $\{ n_{0}<4.26\,n_{s},\, 6.41<n_{1}<11.27\, n_{s}\}$,
$\{ n_{0}<3.15\,n_{s}, \, 6.07<n_{1}<11.64\,n_{s} \}$, for the two cases, respectively.
However, the lower limit of  $\mu_{0}$ and  $n_{0}$ cannot be determined by the astronomical observations.
Meanwhile the distribution of the phase transition chemical potentials can be fitted with the Gaussian distribution.
With the fitting, we find that the most probable phase transition chemical potentials are $\langle\mu_{0}\rangle=1.00\,\textrm{GeV}$, $\langle\mu_{1}\rangle=1.51\,\textrm{GeV}$
if the hadron matter sector does not contain hyperons,
and $\langle\mu_{0}\rangle=0.99\,\textrm{GeV}$, $\langle\mu_{1}\rangle=1.53\,\textrm{GeV}$
if the hadron matter sector involves hyperons.
The corresponding baryon number density is $\{ \langle n_{0} \rangle , \, \langle n_{1}\rangle \}
= \{1.50, \, 7.90\}\,n_{s}$, $\{ 1.43, \, 8.20\}\,n_{s}$, for each of the two kind matters, respectively.
Such obtained results agree with the ones given via effective field theory of QCD very well.

We have also looked over the effect of the maximum mass, the radius, and the tidal deformability
on the phase transition chemical potentials by varying the calibration ranges of the observables.
We find that the upper boundary of the maximum mass affects the lower limit of the $\mu_{1}$,
and the lower boundary of the tidal deformability determines the upper limit of the $\mu_{1}$.
For the astronomical observation values we employed, the radius is the most loosely constrained one.
However, since the radius, the tidal deformability and the maximum mass are correlated,
it is certain that when the range of the radius is narrow enough,
it will also affect the phase transition density obviously.

For now, the phase transition chemical potentials have not yet been constrained to a very small range.
However, we have found the most probable value of the phase transition chemical potentials.
With future detections, the maximum mass, the tidal deformability and the radius of neutron stars
can be measured with higher accuracy.
The range of the phase transition chemical potentials can be constrained to a better range.
Meanwhile, it is also necessary to look for other astronomical observations to constrain
the phase transition condition more strictly.

% Sections that will go in second font

% Acknowledgement
\section{ACKNOWLEDGMENTS}
The work was supported by the National Natural Science Foundation of China under Contracts No.\ 11435001,
No. 11775041 , and the National Key Basic Research Program of China under Contract No.\ 2015CB856900.

%eReferences

\nocite{*}
\bibliographystyle{aipnum-cp}%
\bibliography{sample}%

\end{document}